\documentclass[hyper,12pt]{JHEP3}
\usepackage{amsmath,amsfonts}
\usepackage{graphicx}

\def\cN{\mathcal{N}}
\def\bZ{\mathbb{Z}}
\def\bR{\mathbb{R}}
\def\SU{SU}
\def\SO{SO}
\def\Tr{\mathop{\mathrm{Tr}}}
\def\Im{\mathop{\mathrm{Im}}}
\def\vev#1{\langle#1\rangle}
\def\cF{\mathcal{F}}

\title{Liouville Correlation Functions \\
from Four-dimensional Gauge Theories}

\author{Luis F. Alday, Davide Gaiotto and Yuji Tachikawa \\
School of Natural Sciences, Institute for Advanced Study, \\
Princeton, NJ 08540, USA
\\
{\tt alday,dgaiotto,yujitach@ias.edu} }

\preprint{ }

\abstract{We conjecture an expression for the Liouville theory conformal
blocks and correlation functions on a Riemann surface of genus
$g$ and $n$ punctures as the Nekrasov partition function of a
certain class of ${\cal N}=2$ SCFTs recently defined by one of
the authors. We conduct extensive tests of the conjecture at
genus $0,1$. \\

}

\begin{document}
\section{Introduction}

In \cite{Gaiotto:2009we} it was shown that
a large class of ${\cal N}=2$
superconformal field theories (SCFTs) in four dimensions
arise by compactifying the six-dimensional (2,0) theory of type $A_1$
on a Riemann surface with punctures.
A canonical example is
${\cal N}=4$ $SU(2)$ gauge theory,
which is the compactification of this 6d theory
on a torus whose modulus is the  marginal gauge coupling parameter $\tau$.
The $SL(2,\bZ)$  S-duality of $\cN=4$ theory is geometrically realized
as the modular transformation of the torus.

Each SCFT in the class is thus labeled by two integers $g,n$, which
are the genus and the number of punctures of the Riemann surface.
The parameter space of the theory
coincides with the complex structure moduli space of the
punctured Riemann surface.
Each distinct way to sew such Riemann surface together from
pairs of pants corresponds to a different Lagrangian
description of the theory, and
a Lagrangian description is weakly coupled in
the region of parameter space where the pairs of pants are sewn
together by long tubes. Each tube represents an $SU(2)$ gauge
group. Each pair of pants represents a ``block'' of matter
hypermultiplets. The sewing of the Riemann surface encodes the
detailed structure of the matter representations.

It was also shown in \cite{Gaiotto:2009we} that
each puncture is associated to an $\SU(2)$ flavor symmetry,
which can be used to give  mass to the hypermultiplets.
We have already mentioned that the compactification on
a torus without a puncture gives rise to the $\cN=4$ theory.
Other basic examples are
a sphere with four punctures which gives $\cN=2$ $\SU(2)$ theory with $N_f=4$ flavors,
and a torus with one puncture which gives $\cN=2^*$ $\SU(2)$ theory,
i.e. $\cN=4$  theory deformed by the mass to the  adjoint hypermultiplet.
The former has $\SO(8)$ flavor symmetry which contains as a subgroup
$\SU(2)^4$, each $SU(2)$ factor corresponding to each of the punctures;
the mass parameter of the latter is associated to  the $\SU(2)$ flavor symmetry
acting on the adjoint hypermultiplet.
Thus each puncture is associated with a number,
which is the mass parameter of the $\SU(2)$ flavor symmetry
associated to it.

Basic operations which are familiar from the theory of sewing
Riemann surfaces have a direct translation in the language of
these ${\cal N}=2$ field theories. The basic operations which
relate different sewings of the same Riemann surface are
elementary S-duality transformations, which relate different
Lagrangian descriptions of the same theory. Sewing two Riemann
surfaces together, or adding a handle to a Riemann surface also
map to very natural operations on the corresponding ${\cal
N}=2$ SCFTs: they correspond to gauging the diagonal
subgroup of two $\SU(2)$ flavor symmetries at the two punctures  sewed together.

There is a huge physical and mathematical literature on ${\cal
N}=2$ field theories. It is natural to wonder if this class of
SCFTs may provide a connection with the large literature on
objects defined through the sewing of Riemann surfaces, in
particular the theory of 2d conformal field theories. This
paper is devoted to test a specific realization of this general
idea: the identification of the Nekrasov partition function \cite{Nekrasov:2002qd,Nekrasov:2003rj} of these ${\cal N}=2$ SCFTs and the Liouville theory correlation
functions on the corresponding Riemann surfaces.

The crucial idea is that for each sewing of the Riemann surface
one is given two natural objects: Nekrasov's instanton
computation in the corresponding Lagrangian description of the
theory and the ``Liouville conformal block" defined by a sum
over Virasoro descendants of a primary field in each of the
sewing channels. With a judicious identification of the
parameters on the two sides, we will demonstrate by explicit
examples that the two objects coincide at genus $g=0,1$ for
various $n$. We will also conjecture the general map at higher
genus and number of punctures.

Taking inspiration from Pestun's  computation \cite{Pestun:2007rz} of the
$S^4$ partition function of ${\cal N}=2$ SCFTs we will assemble
together the squared modulus of Nekrasov's instanton partition
function together with tree level and one-loop contributions to
produce an S-duality invariant object, which coincides with the
Liouville correlation function on the corresponding punctured
Riemann surface. We will see that the product of the
Liouville three-point functions \cite{Dorn:1994xn,Zamolodchikov:1995aa,Teschner:1995yf}
neatly recombines into the modulus squared of the one-loop contribution
to Nekrasov's partition function.

The structure of the paper is the following.
We begin in Sec.~\ref{sec:review}
by reviewing the class of four-dimensional $\cN=2$  SCFTs
associated to punctured Riemann surfaces.
Then in Sec.~\ref{sec:blocks}
we formulate the equivalence of Nekrasov's instanton sum associated
to this class of theories and the Liouville conformal blocks.
In Sec.~\ref{sec:correlators} we go on to show that the Liouville correlators
on a sphere or on a torus can be written as an integral of the absolute value
squared of Nekrasov's full partition function, including the classical and the
one-loop part in addition to the instanton part.
In Sec.~\ref{sec:T} we briefly discuss how the Seiberg-Witten curve can be recovered
from the point of view of the Liouville theory; it involves the insertion
of the energy momentum tensor of the 2d CFT. As a byproduct of this analysis we are led to a proposal for the quantum version of the Seiberg-Witten curve.
We conclude with the discussions of future directions in
Sec.~\ref{sec:speculation}.
There are three appendices: App.~\ref{app:liouville} and App.~\ref{app:nekrasov}
collect rudimentary facts about the Liouville theory
and Nekrasov's instanton counting, respectively.
In App.~\ref{app:u1} we propose how to decouple the $U(1)$ part from
Nekrasov's instanton sum of $U(2)$ quiver theories.

\section{Review: a class of four dimensional ${\cal N}=2$ SCFTs}\label{sec:review}

We will denote  as ${\cal T}_{g,n}$
the four-dimensional SCFT  we obtain by compactifying
six-dimensional (2,0) theory of type $A_1$ on a genus-$g$ Riemann
surface with $n$ punctures.
The parameter space of gauge couplings
is the moduli  ${\cal M}_{g,n}$
of the genus-$g$ $n$-punctured Riemann surfaces.
The surface itself is denoted by $C_{g,n}$,
which is sewn from $2g-2+n$ pairs of pants, joined by $3g-3+n$
tubes.

The simplest example is the theory associated to a three punctured-sphere, ${\cal T}_{0,3}$.
${\cal M}_{0,3}$ is a point, and
${\cal T}_{0,3}$ is simply a theory of four free hypermultiplets. Four free hypermultiplets can be represented by eight $\cN=1$ free chiral multiplets
transforming in $\mathbf{2}_a \otimes \mathbf{2}_b \otimes \mathbf{2}_c$
of the flavor symmetry  $SU(2)_a\times SU(2)_b \times SU(2)_c$
which commutes with $\cN=2$  supercharges. Let us denote
the mass parameters associated to the three $SU(2)$ flavor symmetries
as $m_a$, $m_b$ and $m_c$ respectively.
Then the masses of the hypermultiplets  are \begin{equation}
m_a \pm m_b\pm m_c.
\end{equation}

When sewing pairs of pants together, one always gauges a diagonal combination of two such $SU(2)$ flavor symmetry groups. If the groups belong to different pairs of pants, the $SU(2)$ gauge group is coupled to a total of $8$ hypermultiplets, i.e. four fundamental $SU(2)$ matter representations. If the two flavor groups belong to the same set of hypermultiplets, they represent an adjoint plus a singlet of $SU(2)$. All in all, at the end we are left with $n$ residual $SU(2)$ flavor symmetry groups, each associated to a puncture of $C_{g,n}$. In the following we will label the flavor groups as $SU(2)_{a,b,c,\cdots}$ and the corresponding mass parameters as $m^2_{a,b,c,\cdots}$. We will also denote the $SU(2)$ gauge groups as $SU(2)_{1,2,3,\cdots}$. We denote the Coulomb branch order parameters as $a_{i=1,2,3,\cdots}$.
Semi-classically they are the diagonal components of the adjoint scalar, $\phi_i=\mathrm{diag}(a_i,-a_i)$. More precisely they are special coordinates controlling the mass of the BPS particles.
If we keep track of the flavor symmetry groups, the natural gauge coupling parameter space is the moduli space of a Riemann surface with $n$ distinct punctures.

The next simplest example is the ${\cal T}_{0,4}$ theory. A  sphere with four punctures can be assembled from two pairs of pants joined by a tube. All weakly coupled realizations of the theory involve a single  $SU(2)$ gauge group, coupled to a total of four fundamental hypermultiplets: $SU(2)$ $N_f=4$.  This theory is very well studied \cite{Seiberg:1994aj}. It has an overall $SO(8)$ flavor symmetry group,  a marginal gauge coupling $\tau$ and a peculiar S-duality group: it is $SL(2,\bZ)$, but some S-duality transformations
exchange the standard matter fields in the $\mathbf{8}_v$ representation of $SO(8)$ with new
matter fields in $\mathbf{8}_s$ or $\mathbf{8}_c$ representation.

\begin{figure}
\[
 \includegraphics[scale=.5]{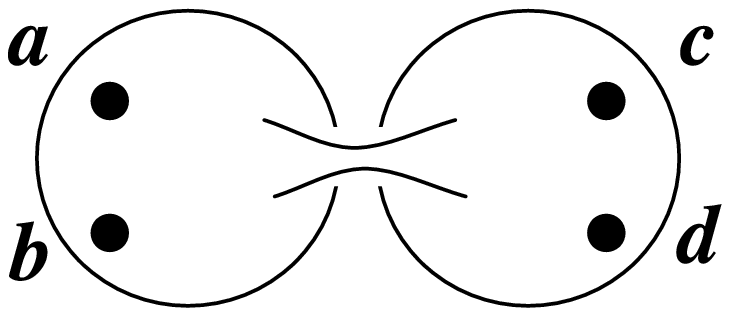}\qquad
 \includegraphics[scale=.5]{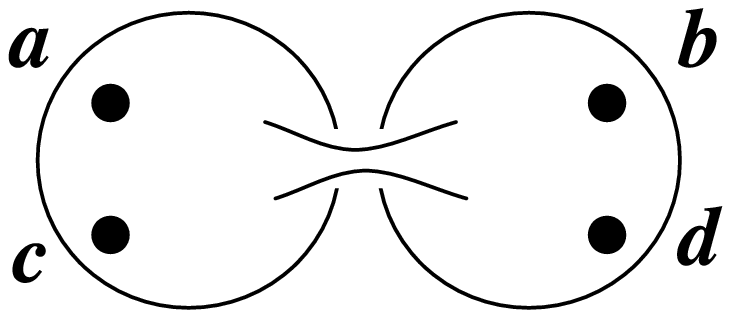}\qquad
 \includegraphics[scale=.5]{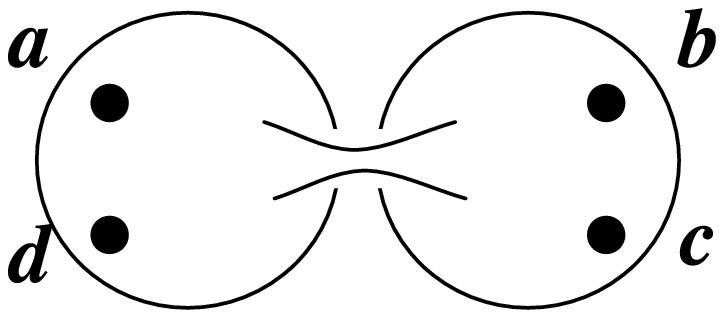}
\]
\caption{Three distinct ways to decompose a four-punctured sphere into
two pairs of pants. They correspond to three distinct weakly-coupled frames
of $SU(2)$ gauge theory with four flavors. \label{fig:nf4}}
\end{figure}

From the point of view of the pants decomposition, it is natural to consider the two
groups of two fundamental hypermultiplets separately, and focus on an $SO(4) \times SO(4) \sim SU(2)_a\times SU(2)_b \times SU(2)_c \times SU(2)_d$  subgroup of the flavor group. Then
\begin{align}
 \mathbf{8}_v \sim (\mathbf{2}_a \otimes \mathbf{2}_b) \oplus (\mathbf{2}_c \otimes \mathbf{2}_d), \\
 \mathbf{8}_s \sim (\mathbf{2}_a \otimes \mathbf{2}_c) \oplus (\mathbf{2}_b \otimes \mathbf{2}_d), \\
 \mathbf{8}_c \sim (\mathbf{2}_a \otimes \mathbf{2}_d) \oplus (\mathbf{2}_c \otimes \mathbf{2}_b).
\end{align}
We recognize the three possible ways to decompose the four-punctured sphere in two pairs of pants,
distributing the four punctures $a,b,c,d$ in various ways among the pants.
They correspond to distinct weakly-coupled limits of the same theory,
see Figure~\ref{fig:nf4}.
Let $m_a$ be the mass parameter associated to $SU(2)_a$, etc.
Then the mass eigenvalues of the four hypermultiplets in $\mathbf{8}_v$ is
\begin{equation}
m_a\pm m_b, \qquad m_c \pm m_d.
\end{equation}

There are two natural ways to parameterize the modulus of the sphere
with four punctures. One is to take the cross ratio $q$ of the four punctures.
$q$ lives in $\mathbb{CP}^1\setminus \{0,1,\infty \}$.
The other is to take the double cover of the sphere
and to take the modulus $\tau_\text{IR}$
of the resulting torus, which is the Seiberg-Witten curve when all masses are set to zero.
$\tau_\text{IR}$ parameterizes an upper half plane which
can be seen as the universal cover of the punctured sphere parameterized by $q$.

In the early literature on the subject \cite{Seiberg:1994aj}
$\tau_\text{IR}$ was identified
with $\tau_\text{UV}$. This proposal was invalidated by explicit instanton computations \cite{Dorey:1996bn}.
As we will review in Appendix~\ref{sec:renormalization},
the cross ratio is actually given by the exponential of the UV coupling,
$q=\exp(2\pi i \tau_\text{UV})$; this precise relation was first noticed by \cite{Grimm:2007tm}.

Note that $q$ is also the sewing parameter for the four-punctured sphere.
This result is much more natural in our general setup: the universal cover of
the space of the marginal couplings, that is the moduli space ${\cal M}_{g,n}$
of $n$ punctures on a genus-$g$ Riemann surface in general is
very intricate, and bears no obvious resemblance to the product of
upper half planes parameterized by the gauge couplings $\tau_i$ of the $SU(2)_i$ gauge groups.
It is also distinct from the space of IR gauge couplings, even for zero masses, which actually
depend on the Coulomb branch parameters as well.
On the other hand the set of sewing parameters $q_i$
could be easily matched to the UV gauge couplings as
$q_i = \exp(2 \pi i \tau_{i,\text{UV}})$.

Another simple example is ${\cal T}_{1,1}$, i.e. a torus with one puncture.
It can be assembled
from a pair of pants by gluing together two legs. Hence ${\cal T}_{1,1}$ coincides with an $SU(2)$ gauge theory with an adjoint hypermultiplet (and one extra free hypermultiplet), i.e. ${\cal N}=2^*$ $SU(2)$ gauge theory. The gauge coupling $\tau$
naturally parameterizes the complex structure of the torus, and the sewing parameter is indeed $q =  \exp(2 \pi i \tau)$ again. In this case there is no distinction between
the UV and the IR couplings when the adjoint mass is zero,
because the theory is then $\cN=4$.

\begin{figure}
\[
 \includegraphics[scale=.4]{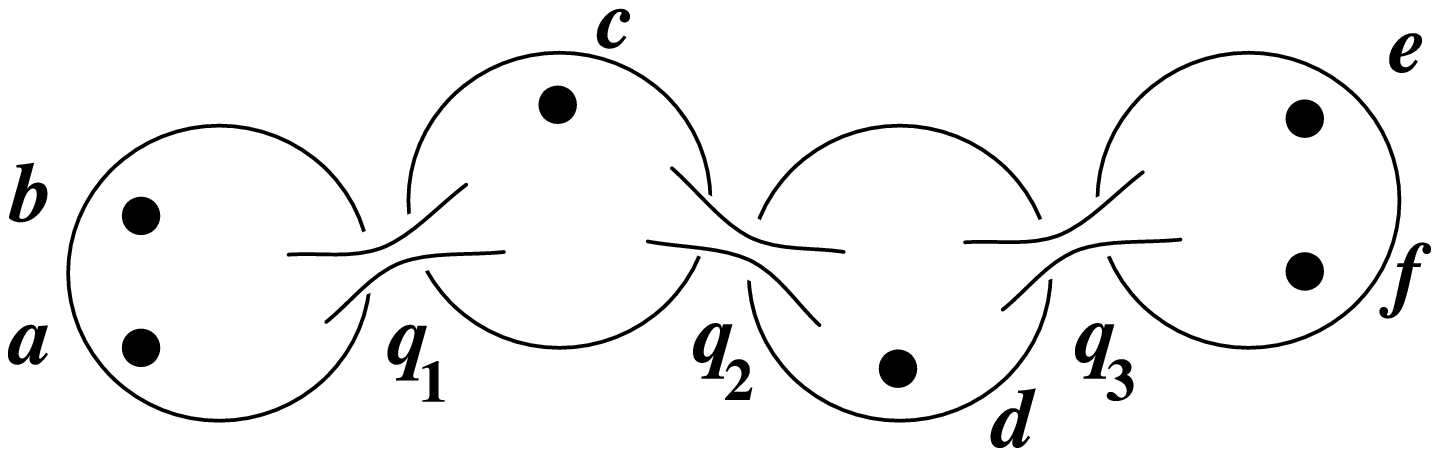}\qquad
 \includegraphics[scale=.4]{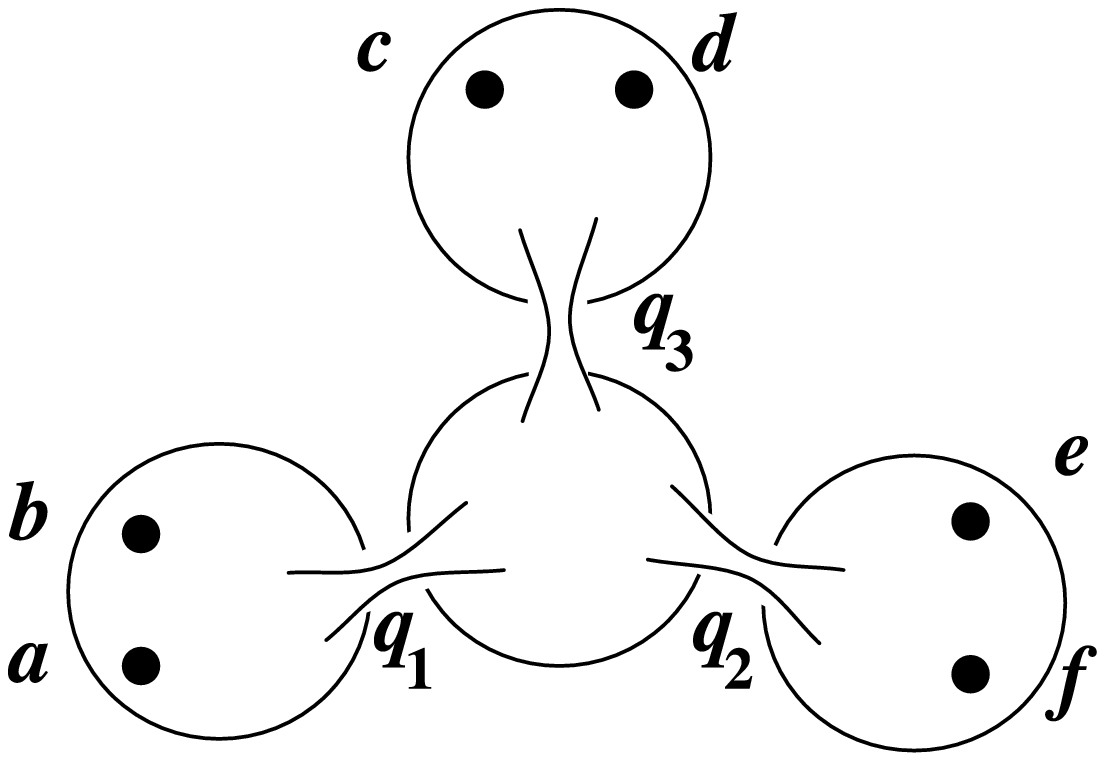}
\]
\caption{Two examples to sew  a six-punctured sphere from
four pairs of pants. Left: a standard linear quiver theory.
Right: a generalized quiver theory, where
three $SU(2)$ gauge groups couple to four hypermultiplets,
denoted by the three-punctured sphere at the center. \label{fig:T06}}
\end{figure}

Next consider a more complex example: a sphere with six punctures, 
which can be sewn from four pairs
of pants in various ways. Two possibilities are shown in Fig.~\ref{fig:T06}.
Both give rise to theories with the gauge group
$\SU(2)_1\times \SU(2)_2\times \SU(2)_3$,
one for each of  the three thin necks with sewing parameter $q_{1,2,3}$.
We again denote as $m_a$ the mass parameter for the $\SU(2)_a$, etc.
Then, the hypermultiplet content of the theory on the left
is two fundamentals of $\SU(2)_1$ with masses $m_a\pm m_b$,
one bifundamental of $\SU(2)_1\times \SU(2)_2$ with mass $m_c$,
another bifundamental of $\SU(2)_2\times \SU(2)_3$ with mass $m_d$,
and two fundamentals of $\SU(2)_1$ with masses $m_e\pm m_f$.
The matter content of the theory on the right is more exotic:
each of $\SU(2)_{1,2,3}$ has two fundamentals with masses
$m_a\pm m_b$, $m_c\pm m_d$ and $m_e\pm m_f$,  respectively,
and the three $\SU(2)_{1,2,3}$  all couple a single set of eight $\cN=1$ chiral multiplets
transforming in $\mathbf{2}_1\otimes\mathbf{2}_2\otimes\mathbf{2}_3$.
This is the simplest example of a ``generalized quiver'' introduced
by one of the authors in \cite{Gaiotto:2009we}.

In this paper, we will focus on examples where each block of hypermultiplets is either a bifundamental hypermultiplet, or a pair of fundamental hypermultiplets. Hence the ``generalized quivers'' will simply be linear quivers of $SU(2)$ gauge groups realizing ${\cal T}_{0,n}$, or necklace quivers
realizing ${\cal T}_{1,n}$.

\section{Instanton sums and the conformal blocks}\label{sec:blocks}

As we reviewed in the previous section,
we can associate an $\cN=2$ SCFT to a punctured Riemann surface,
and we have a Lagrangian description of this SCFT
for each possible way to compose the Riemann surface out of
three-punctured spheres.

There are several quantities of interest which can be computed, given the
Lagrangian of a four-dimensional field theory. When we combine them
with the Lagrangian description for different sewings of  a Riemann surface,
we have a function on the space of the sewings of the Riemann surface.

We will use below the Nekrasov partition function \cite{Nekrasov:2002qd,Nekrasov:2003rj} of the
Lagrangian field theory to produce an interesting function; we will see
that the resulting function on the space of the sewing
is just the standard conformal block of the Virasoro algebra.
In order to express our proposal, we need to
 review briefly what the Nekrasov partition function is.

\subsection{Nekrasov's partition function}
For a given  four-dimensional $\cN=2$  field theory,
Nekrasov considered a deformation of its Lagrangian
by two deformation parameters $\epsilon_{1,2}$ parameterizing
the $SO(4)$ rotation  of the spacetime $\bR^4$.
This breaks the translational symmetry of the system.
The partition function is just a number, which depends meromorphically on the
coupling constants $\tau$, vevs $a$ of the adjoint scalars in the vector multiplets,
and hypermultiplet masses $m$.  The Nekrasov partition function consists of three parts:
the classical, the one-loop, and the instanton parts:
\begin{equation}
Z_\text{full}(\tau,a,m;\epsilon_i) =
Z_\text{classical}
Z_\text{1-loop}
Z_\text{instanton}.
\end{equation}
It has the important property that it gives the prepotential of the theory
in the limit $\epsilon_{1,2}\to 0$ \begin{equation}
F(\tau,a,m) = \lim_{\hbar\to 0} \hbar^2 \log Z_\text{full} (\tau,a,m;\hbar,-\hbar).
\end{equation}
This limit was evaluated for a number of $\cN=2$ theories,
and reproduced the prepotential as determined by
the Seiberg-Witten curve.

The path integral of the Lagrangian deformed by $\epsilon_{1,2}$
localizes to instanton configurations sitting at the origin of the spacetime $\bR^4$.
For $U(N)$ gauge group, such instantons are labeled
by an $N$-tuple of Young tableaux $\vec Y=(Y_1,\ldots,Y_N)$.
The instanton number of the configuration is
given by the number of the boxes $|\vec Y|$.

Then the instanton part of the Nekrasov partition function
is the summation over the Young tableaux, whose summand
is the product of factors corresponding to the field content of the Lagrangian.
As an example, we give the expression for $U(2)$ gauge theory
with one fundamental hypermultiplet:
\begin{equation}
Z^{U(2), N_f=1}_\text{inst}(q,\vec a,m)=\sum_{\vec Y}
q^{|\vec Y|}
z_\text{vector}(\vec a,\vec Y)
z_\text{fund}(\vec a,\vec Y,m), \label{nf=1}
\end{equation}
where $\vec a=(a_1,a_2)$ is  the vev of the adjoint scalar
and $m$ is the mass of the hypermultiplet.
The explicit form of $z_\text{vector}$ and $z_\text{fund}$
can be found in Appendix~\ref{app:nekrasov}.

There are two subtleties of Nekrasov's formulation which will
complicate our investigation.
One is that the deformation by $\epsilon_{i}$
treats hypermultiplets in complex conjugate representations $R$ and $R^*$
differently, so that
the contribution of a hypermultiplet in the representation $R$ with mass $m$
is equivalent to that of another hypermultiplet in $R^*$
with mass $\epsilon_1+\epsilon_2-m$:
\begin{equation}
z_{R}(m) = z_{R^*}(\epsilon_1+\epsilon_2-m).
\end{equation}
The other is that we will use Nekrasov's partition function for
$U(2)$ quiver theories, not for $SU(2)$ quiver theories.
For $U(2)$, the doublet and the anti-doublet are two distinct representations.
Therefore, the expression \eqref{nf=1} above does not have the
symmetry under $m \to \epsilon_1+\epsilon_2-m$
which should be there for $SU(2)$ gauge theory.

We propose to remedy this situation by decoupling the contribution
of $U(1)$ gauge fields from the instanton partition function
so that the symmetry under $m\to \epsilon_1+\epsilon_2 -m$ is recovered.
We find it rather nontrivial that the decoupling can be consistently performed at all.

\subsection{Sphere with four punctures}

Here and in the following, we will deal with superconformal theories. The deformation parameters $\epsilon_i$,
the vevs $a$ and the masses $m_i$ all have mass dimension one.
We choose to fix the scale by setting \begin{equation}
\epsilon_1=b,\qquad \epsilon_2=1/b.
\end{equation}
We also define \begin{equation}
Q=\epsilon_1+\epsilon_2 = b+1/b.
\end{equation}

Let us consider the simplest case,
the six-dimensional (2,0) theory of type $A_1$
compactified on a sphere with four punctures.
The manifest flavor symmetry in this description is $\SU(2)^4$,
one $\SU(2)$ factor for each puncture.
At low energy, this becomes $\cN=2$ $\SU(2)$ gauge theory with $N_f=4$ flavors
and the flavor symmetry enhances to $\SO(8)$.

We write down
Nekrasov's instanton partition function for $U(2)$ theory with $N_f=4$
flavors instead, which is given by the formula
\begin{multline}
Z_\text{inst}^{U(2),N_f=4}
=\sum_{\vec Y} q^{|\vec Y|} z_\text{vector}(\vec a,\vec Y) \\
z_\text{antifund}(\vec a,\vec Y,\mu_1)
z_\text{antifund}(\vec a,\vec Y,\mu_2)
z_\text{fund}(\vec a,\vec Y,\mu_3)
z_\text{fund}(\vec a,\vec Y,\mu_4).\label{U(2)Nek}
\end{multline}
Here $\vec a=(a_1,a_2)$ is the adjoint vev of the $U(2)$ gauge multiplet,
$\mu_{1,2}$ are the masses of two hypermultiplets in the
anti-fundamental, and $\mu_{3,4}$ are those of the fundamentals.
Explicit expressions for the contributions
$z_\text{vector,fund,antifund}$
can be found in Appendix~\ref{app:nekrasov}.

Manifest flavor symmetries are now $U(2)_1\times U(2)_2$,
acting on $\mu_{1,2}$ and
$\mu_{3,4}$, respectively. We redefine them as follows:\begin{align}
\mu_1& =m_0+\tilde m_0,  &
\mu_2& =m_0- \tilde m_0,  &
\mu_3& =m_1+\tilde m_1,  &
\mu_4& =m_1- \tilde m_1.
\end{align}
$m_i$ is the mass parameter associated to $U(1)_i\subset U(2)_i$,
and $\tilde m_i$ is the one associated  to $\SU(2)_i\subset U(2)_i$.

Let us stress that the formula \eqref{U(2)Nek} is for $U(2)$ gauge group;
one expects that by decoupling the $U(1)$ part of the gauge group
the manifest flavor symmetry $U(2)_i=SU(2)_i\times U(1)_i$ would enhance
to $SU(2)_i \times SU(2)_{\tilde i}$.
To do that, one needs to set $\vec a=(a_1,a_2)=(a,-a)$,
and also to eliminate the contribution from the $U(1)$ gauge multiplet.
Without further ado,  we propose  how to decouple the  $U(1)$ part:
\begin{equation}
Z_\text{inst}^{U(2),N_f=4}(a,m_0,\tilde m_0,m_1,\tilde m_1)
= (1-q)^{2m_0(Q-m_1)}
\cF_{\alpha_0}{}^{m_0}{}_\alpha{}^{m_1}{}_{\alpha_1}  (q)
\end{equation}
where the relation between $\alpha,\alpha_i$ and the mass parameters is given by
\begin{align}
\alpha&=Q/2 +a ,&\alpha_0 &= Q/2+\tilde m_0, &\alpha_1&= Q/2+ \tilde m_1.
\end{align}
By explicit calculation, one can check that
 $\cF_{\alpha_0}{}^{m_0}{}_\alpha{}^{m_1}{}_{\alpha_1}  (q) $
is a function invariant under individual flips of $\alpha$, $\alpha_i$, and $m_i$:
\begin{equation}
\alpha\to Q-\alpha, \qquad
\alpha_i\to Q-\alpha_i,\qquad
m_i\to Q-m_i.
\end{equation}

We identify the flip of $\alpha$ as the action of the Weyl group
of the $\SU(2)$ gauge symmetry,
and the flips of $\alpha_i$, $m_i$ as that
of the four $SU(2)$ flavor symmetries.
The prefactor $(1-q)^{2m_0(Q-m_1)}$
is {\em not} invariant under the flip of $m_{0,1}$,
but is independent of $a$, which is the expected property for the contribution of the $U(1)$ gauge field.

\begin{figure}
\[
\begin{array}{l@{\qquad}l}
a) \quad \cF_{\alpha_0}{}^{m_0}{}_\alpha{}^{m_1}{}_{\alpha_1} &
b) \quad \cF{}_\alpha{}^m\\[.5em]
\includegraphics[scale=.5]{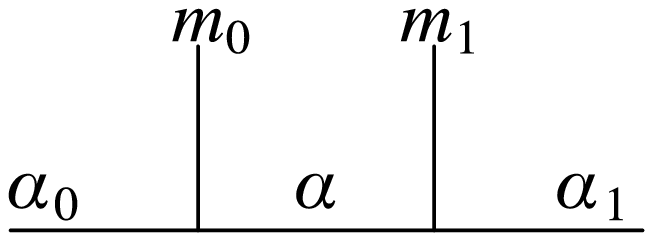} &
\includegraphics[scale=.5]{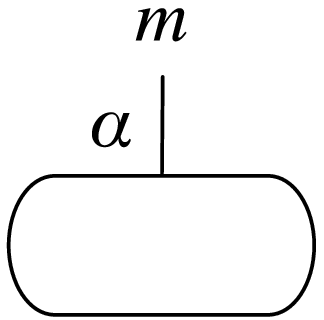}  \\[2em]
c) \quad \cF_{\alpha_0}{}^{m_0}{}_{\alpha_1}{}^{m_1}\cdots
_{\alpha_n}{}^{m_{n}}{}_{\alpha_{n+1}} &
d) \quad \cF{}_{\alpha_1}{}^{m_1}\cdots
_{\alpha_n}{}^{m_n} \\[.5em]
 \includegraphics[scale=.5]{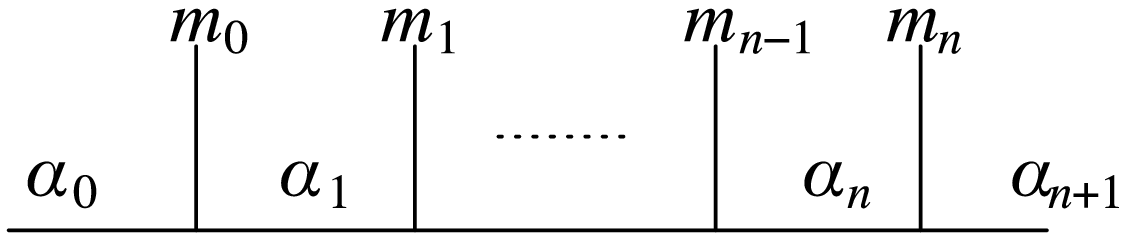} &
 \includegraphics[scale=.5]{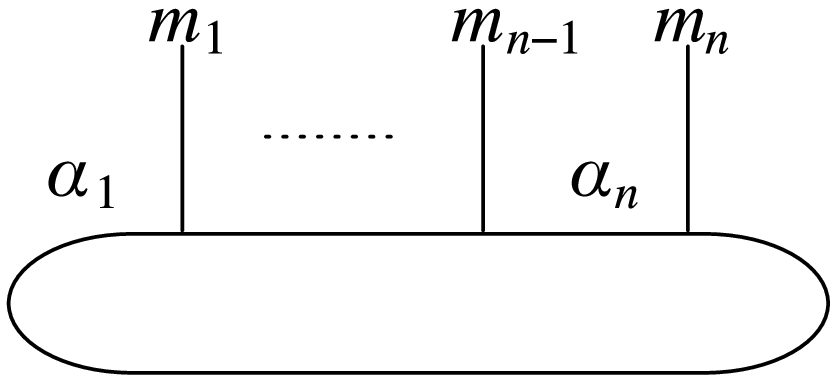}  
\end{array}
\]
\caption{Placement of labels  of the conformal blocks we use. \label{block}}
\end{figure}

Now we come to a surprising observation: explicit calculation\footnote{We checked this statement up to order $q^{11}$ in the instanton expansion.}
tells us  that $\cF_{\alpha_0}{}^{m_0}{}_\alpha{}^{m_1}{}_{\alpha_1}  (q)$
is exactly the conformal block of the Virasoro algebra
with  central charge $c=1+6Q^2$
for four operators of dimensions $\Delta_{1,2,3,4}$
inserted at $\infty,1,q,0$, respectively and with an intermediate state
in the $s$-channel whose dimension is $\Delta$, see Figure~\ref{block} a).
Here \begin{align}
\Delta&=\alpha(Q-\alpha),&
\Delta_1&=\alpha_0(Q-\alpha_0),&
\Delta_2&=m_0(Q-m_0),\\
&&\Delta_3&=m_1(Q-m_1),&
\Delta_4&=\alpha_1(Q-\alpha_1).
\end{align}

Note that $\cF_{\alpha_0}{}^{m_0}{}_\alpha{}^{m_1}{}_{\alpha_1}  (q)$
does not have the symmetry under the permutation
of $\alpha_{0,1}$ and $m_{0,1}$ keeping $q$ fixed.
In other words, it is not invariant under $SO(8)$ keeping $q$ fixed.
It should not come as a surprise, because
even the Seiberg-Witten curve of this theory
in the formalism reviewed in Sec.~\ref{sec:review}
does not have the manifest $SO(8)$ symmetry.

\subsection{Torus with one puncture}
The second simplest example is the $\cN=2^*$ theory,
i.e.~$\cN=4$ $\SU(2)$ theory deformed by a mass term
for the adjoint hypermultiplet.
For the $\cN=2^*$ $U(2)$ theory, Nekrasov's instanton partition function is
\begin{equation}
Z_\text{inst}^{\cN=2^*,U(2)}
=\sum_{\vec Y} q^{|\vec Y|} z_\text{vector}(\vec a,\vec Y)
z_\text{adj}(\vec a,\vec Y,m).
\end{equation}
This is invariant under the flip $m\to Q-m$.
Encouraged by our observation in the previous subsection,
we might hope that we would get a conformal block
by setting $\vec a=(a_1,a_2)=(a,-a)$
and  splitting off the contribution from the $U(1)$ vector multiplet.
Indeed, by some trial and error one finds \begin{equation}
Z_\text{inst}^{\cN=2^*,U(2)}(q,a,m)=\left[\prod_{i=1}^\infty (1-q^i)\right]^{-1+2m(Q-m)} \cF{}_\alpha{}^m(q),
\end{equation} where $\cF{}_\alpha{}^m(q)$ is
the conformal block of the Virasoro algebra of central charge $c=1+6Q^2$
on a torus whose modulus is $q$, with one operator of dimension
$\Delta_1=m(Q-m)$ inserted and a primary of  dimension
$\Delta=\alpha(Q-\alpha)$ in the intermediate channel \footnote{We have checked this up to order $q^7$ in the instanton expansion.};
here $\alpha=Q/2+a$ as before. See Figure~\ref{block} b).

\subsection{Sphere with multiple punctures}
Now let us move on to more complicated cases.
A sphere with $n+3$ punctures has a weakly-coupled
description as  a linear quiver gauge theory
with $n$ $\SU(2)_i$ gauge groups, $i=1,\ldots,n$, with coupling constants $q_i$
and the vevs $a_i$.
The hypermultiplets are
two antifundamentals of mass $\mu_{1,2}$ on $\SU(2)_1$,
one bifundamental of mass $m_i$ between $\SU(2)_{i}$ and $\SU(2)_{i+1}$,
and two fundamentals of mass $\mu_{3,4}$ on $\SU(2)_n$.

It is straightforward to write down Nekrasov's instanton sum
for the quivers of $U(2)$ gauge groups.
Again it is natural to rewrite the masses of (anti)-fundamentals
as \begin{equation}
\mu_1=m_0+\tilde m_0, \quad
\mu_2=m_0-\tilde m_0,\quad
\mu_3=m_{n}+\tilde m_1,\quad
\mu_4=m_{n}-\tilde m_1.
\end{equation}

We propose to decouple the $U(1)$ factor in the following way:
\begin{multline}
Z_\text{inst}^\text{ $U(2)$ linear quiver}(q_i;a_i;m_i;\tilde m_i)\\
= Z^\text{$U(1)$ linear}(q_i;m_i) \cF_{\alpha_0}{}^{m_0}{}_{\alpha_1}{}^{m_1}\cdots
_{\alpha_n}{}^{m_{n}}{}_{\alpha_{n+1}}(q_1,q_2,\ldots,q_n)
\end{multline}
where \begin{equation}
\alpha_0=Q/2+\tilde m_0,\quad
\alpha_i=Q/2+a_i,\quad
\alpha_{n+1}=Q/2+\tilde m_1.
\end{equation}
The $U(1)$ factor $Z^\text{$U(1)$ linear}(q_i;m_i)$ is detailed in App.~\ref{app:u1};
this is a function only of the coupling constants $q_i$
and the masses $m_i$, and is {\em not} symmetric under $m_i\to Q-m_i$.

$\cF_{\alpha_0}{}^{m_0}{}_{\alpha_1}{}^{m_1}\cdots
_{\alpha_n}{}^{m_{n}}{}_{\alpha_{n+1}}(q_1,q_2,\ldots,q_n)$
can then be seen to be
 the conformal block of Virasoro algebra with central charge $c=1+6Q^2$
for  a sphere with $n+3$ punctures at
\begin{equation}
 \infty ,\  1 ,\   q_1 ,\   q_1q_2 ,\  \ldots,  q_1q_2\cdots q_n ,\   0 .
\end{equation}
The dimensions of the operators at the punctures are
\begin{equation}
\alpha_0(Q-\alpha_0),\
 m_0(Q-m_0),\
 \ldots,\
 m_{n}(Q-m_{n}),\
\alpha_{n+1}(Q-\alpha_{n+1})
\end{equation}
 respectively,
and that of the operator in the $i$-th intermediate channel
is $\alpha_i(Q-\alpha_i)$, see Figure~\ref{block} c).

\subsection{Torus with multiple punctures}

Let us next consider a torus with $n$ points.
It  has a weakly-coupled
description as  a necklace quiver gauge theory
with $n$ $\SU(2)_i$ gauge groups, $i=1,\ldots,n$, with coupling constants $q_i$
and the vevs $a_i$.
The hypermultiplets are
one bifundamental of mass $m_i$ between each consecutive
pair of gauge groups $\SU(2)_{i}$ and $\SU(2)_{i+1}$;
we identify $\SU(2)_{n+1}$ with $\SU(2)_1$.

It is again straightforward to write down Nekrasov's instanton sum
for the quivers of $U(2)$ gauge groups.
We propose to decouple the $U(1)$ factor in the following way:
\begin{multline}
Z_\text{inst}^\text{ $U(2)$ necklace quiver}(q_i;a_i;m_i)\\
= Z^\text{$U(1)$ necklace}(q_i;m_i) \cF{}_{\alpha_1}{}^{m_1}\cdots
_{\alpha_n}{}^{m_n}(q_1,q_2,\ldots,q_n)
\end{multline}
where $\alpha_i=Q/2+a_i$.
The $U(1)$ factor is again detailed in App.~\ref{app:u1};
this is a function only of the coupling constants $q_i$
and the masses $m_i$, and is {\em not} symmetric under $m_i\to Q-m_i$ in general.

$\cF{}_{\alpha_1}{}^{m_1}\cdots
_{\alpha_n}{}^{m_n}(q_1,q_2,\ldots,q_n)$
can then be seen to be
 the conformal block of Virasoro algebra with central charge $c=1+6Q^2$
for a torus with modulus $q_1q_2\cdots q_n$ with
punctures at
\begin{equation}
1,\  q_1 ,\   q_1q_2 ,\   \ldots,\  q_1q_2\cdots q_{n-1}.
\end{equation}
The dimension of the operator at the $i$-th puncture is
$ m_i(Q-m_i)$,
and that of the operator in the $i$-th intermediate channel
is $\alpha_i(Q-\alpha_i)$, see Figure~\ref{block} d).

\section{Liouville correlators}\label{sec:correlators}
In the last section we presented a concrete way to decouple
the $U(1)$ part from Nekrasov's instanton partition function for
$U(2)$ quivers.  This resulted in the identification
of the $SU(2)$ part of the instanton sum with the Virasoro conformal block
with central charge $c=1+6Q^2$.

 Conformal blocks, combined with three-point functions,
give multi-point correlators of concrete CFTs.
This then begs a natural question: is there
something we need to combine with
the instanton part of Nekrasov's partition function,
to form another physical quantity?
Indeed there is such a thing, which is
the one-loop part of Nekrasov's partition function.

From the point of view of the low-energy theory,
splitting the prepotential into the one-loop and the instanton parts is rather artificial
in that the split depends on the electromagnetic frame one is interested in.
This is analogous to the situation with the multi-point correlators of a 2d CFT:
decomposition of the correlators into the products of three-point functions
and conformal blocks depends on the channel one is interested in.

We will show below that the one-loop factor precisely reproduces
the product of  the DOZZ three-point functions \cite{Dorn:1994xn,Zamolodchikov:1995aa,Teschner:1995yf} of the Liouville theory.
In other words, the absolute value squared of the full Nekrasov partition function,
integrated over the vevs $a_i$, is a {\em  Liouville correlator.}

\subsection{Sphere with four punctures}
Without further ado, let us consider the four-point function of the Liouville theory
on a sphere,
\begin{multline}
\vev{V_{\alpha_0}(\infty)V_{m_0}(1)V_{m_1}(q)V_{\alpha_1}(0)}\\
=\int \frac{d\alpha}{2\pi}  C(\alpha_0^*,m_0,\alpha)
C(\alpha^*,m_1,\alpha_1)
 \left| q^{\Delta_\alpha -\Delta_{m_1}-\Delta_{\alpha_1}}
\cF_{\alpha_0}{}^{m_0}{}_{\alpha}{}^{m_1}{}_{\alpha_1}  (q)\right| ^2.
\end{multline}
Here and in the following,  $V_\alpha(z)=e^{2\alpha\phi(z)}$ is the
Liouville exponential, and
we take all $\alpha$, $\alpha_i$, $m_i$ to be
$\in Q/2 + i \bR$; the integral over $\alpha$ is along this line.
We also use the notation $\alpha=Q/2+a$, $\alpha_i=Q/2+\tilde m_i$,
$m_i=Q/2+\hat m_i$.
The three-point function $C(\alpha_1,\alpha_2,\alpha_3)$ is given by the
DOZZ formula, see Appendix~\ref{app:liouville}. Using formulae
collected there, one can massage the right hand side into the form \begin{equation}
=  f(\alpha_0^*)f(m_0)f(m_1)f(\alpha_1)
\left| q^{Q^2/4 -\Delta_{m_1}-\Delta_{\alpha_1}} \right|^2
 \int {a^2 da}
\left|Z_{\alpha_0}{}^{m_0}{}_{\alpha}{}^{m_1}{}_{\alpha_1}  (q)\right|^2 \label{OSV}
\end{equation} up to a constant which only  depends on $b$.
Here \begin{equation}
f(\alpha)=\left[\pi \mu \gamma(b^2)b^{2-2b^2}\right]^{-\alpha/b}\Upsilon(2\alpha)
\end{equation}
and  \begin{multline}
Z_{\alpha_0}{}^{m_0}{}_{\alpha}{}^{m_1}{}_{\alpha_1}   (q)
=\\
q^{-a^2} \frac{\prod \Gamma_2( \hat m_0 \pm  \tilde m_0 \pm a  +Q/2)
 \prod\Gamma_2(\hat m_1 \pm  \tilde m_1 \pm a +Q/2)}
{\Gamma_2(2a+b)\Gamma_2(2a+1/b)}
\cF_{\alpha_0}{}^{m_0}{}_{\alpha}{}^{m_1}{}_{\alpha_1}   (q).
\end{multline}
In the last expression each product is over the four choices of signs.
Using the formula for the one-loop factors collected in Appendix~\ref{app:nekrasov},
\footnote{As will be elaborated there, our one-loop factor for the vector multiplet
is slightly different from that in \cite{Nekrasov:2003rj},
but agrees with that in \cite{Hollowood:2003cv,Iqbal:2007ii}. }
we find it is equal to \begin{multline}
Z_{\alpha_0}{}^{m_0}{}_{\alpha}{}^{m_1}{}_{\alpha_1}   (q)
=
q^{-a^2} \times
z_\text{vector}^\text{1-loop}(a)
z_\text{antifund}^\text{1-loop}(a,\mu_1)
z_\text{antifund}^\text{1-loop}(a,\mu_2) \\
\times
z_\text{fund}^\text{1-loop}(a,\mu_3)
z_\text{fund}^\text{1-loop}(a,\mu_4)
\cF_{\alpha_0}{}^{m_0}{}_{\alpha}{}^{m_1}{}_{\alpha_1}   (q)
\end{multline}
where \begin{equation}
\mu_1=m_0+\tilde m_0, \quad
\mu_2=m_0-\tilde m_0, \quad
 \mu_3=m_1+\tilde m_1, \quad
\mu_4=m_1-\tilde m_1.
\end{equation}

We identified in the previous section
$\cF _{\alpha_0}{}^{m_0}{}_{\alpha}{}^{m_1}{}_{\alpha_1}   (q)$
as Nekrasov's instanton partition function for $\SU(2)$ theory
with four flavors with masses $\mu_{1,2}$ and $ \mu_{3,4}$.
One can easily see that  $q^{-a^2}$ gives the exponential of the
classical prepotential $(2\pi i)\tau_\text{UV} a^2$,
and the product of the one-loop factors is exactly the one
for this gauge theory. Thus,
$Z _{\alpha_0}{}^{m_0}{}_{\alpha}{}^{m_1}{}_{\alpha_1}   (q)$
is precisely Nekrasov's full partition function of
$\SU(2)$ gauge theory with four flavors.
In the integral \eqref{OSV}, the absolute value squared of this
partition function  is integrated over the natural measure $a^2da$ on the
Cartan subalgebra of $\SU(2)$, including the Vandermonde determinant.
Therefore, we have come to a surprising conclusion that
{\em Nekrasov's full partition function, integrated over the vev 
with the natural measure,
is the Liouville four-point function on the sphere.}

This integral, from the gauge theory point of view,
has appeared in \cite{Pestun:2007rz}
when $b=1/b=1$. There, the integral \eqref{OSV}
without the prefactor, i.e. \begin{equation}
 \int {a^2 da}
\left|Z_{\alpha_0}{}^{m_0}{}_{\alpha}{}^{m_1}{}_{\alpha_1}  (q)\right|^2
\end{equation} appeared as the partition function of the 4d SCFT
on $S^4$.

The Liouville four-point function, as constructed from the DOZZ three-point functions
and the conformal blocks, has been proved in \cite{Ponsot:1999uf,Teschner:2003en} 
to be crossing symmetric. 
Therefore, we find that the 
absolute-value squared of Nekrasov's partition function, integrated over the vev $a$,
is indeed an S-duality invariant object.

\subsection{Torus with one puncture}
We can perform the same procedure
on the one-point function of the Liouville theory on a torus:
\begin{align}
\vev{V_m}_q &= \int \frac{d\alpha}{2\pi}
C(\alpha^*,m,\alpha)  |q^{\Delta_\alpha} \cF{}_\alpha{}^{m} (q)|^2  \\
&= c' f(m) \int a^2 da |Z{}_\alpha{}^m(q)|^2 \label{1-point-on-torus}
\end{align} where \begin{equation}
Z{}_\alpha{}^m(q) = q^{-a^2} z_\text{vector}^\text{1-loop} (a)
z_\text{adj}^\text{1-loop} (a,m) \cF{}_\alpha{}^m(q).
\end{equation}
We have identified in the previous section
$\cF{}_\alpha{}^m(q)$ as the instanton part of
Nekrasov's partition function of $\cN=2^*$ $\SU(2)$  gauge theory.
The DOZZ formula gave us precisely the one-loop factors
for the vector multiplet and the adjoint hypermultiplet, and thus
$Z{}_\alpha{}^m(q)$ is Nekrasov's full partition function of this gauge theory.
Therefore, we find that the one-point function of the Liouville theory on a torus
is the absolute value squared of Nekrasov's full partition function,
integrated over $a$ with the natural measure.

The modular invariance of the torus one-point function of the Liouville theory,
as constructed from the DOZZ three-point function and the conformal block,
has not been fully demonstrated yet, but presumably it can be shown using 
the equivalence of the Liouville theory to the quantum Teichm\"uller theory \cite{Teschner:2003at,Teschner:2005bz,Teschner:2008qh}.
It would be worthwhile to prove the modular invariance of \eqref{1-point-on-torus}\footnote{This was achieved in \cite{Hadasz:2009sw} in November 2009.},
which then implies the S-duality of the $\cN=2^*$ theory.

The one-loop factor cancels when $m=0$ and Eq.\eqref{1-point-on-torus}
reproduces the standard torus amplitude of the Liouville theory. 
This limit corresponds to the $\cN=4$ $SU(2)$ theory.\footnote{
The partition function was  calculated in \cite{Pestun:2007rz}. 
There, the one-loop factor of the vector multiplet used was slightly different from
ours, as we will explain in more detail in Appendix~\ref{sec:one-loop}. This difference 
produces extra powers of $\Im \tau$, effectively replacing 
his $|\eta(\tau)|^2$ by $\sqrt{\Im \tau}| \eta(\tau) |^2$.  This makes the partition function modular invariant. See also \cite{Nanopoulos:2009au}.}

\subsection{General proposal}
The generalization of the analysis above to
multiple points on a sphere and on a torus is now immediate.
A sphere with $n+3$ punctures corresponds to a linear quiver
of $n$ $\SU(2)$ gauge groups.
Then we have the relation \begin{multline}
\vev{V_{\alpha_0}(\infty)V_{m_0}(1)V_{m_1}(q_1)
\cdots V_{m_{n}}(q_1\cdots q_n) V_{\alpha_{n+1}}(0)}=\\
c f(\alpha_0)f(\alpha_{n+1})\prod  f(m_i)\int \prod (a_i^2da_i)
\left|Z_{\alpha_0}{}^{m_0}{}_{\alpha_1}{}^{m_1}\cdots
_{\alpha_n}{}^{m_{n}}{}_{\alpha_{n+1}}(q_i) \right|^2
\end{multline} where
$Z_{\alpha_0}{}^{m_0}{}_{\alpha_1}{}^{m_1}\cdots
_{\alpha_n}{}^{m_{n}}{}_{\alpha_{n+1}}(q_i)$ is the
Nekrasov's full partition function for this $\SU(2)$ quiver gauge theory,
i.e.~$\cF_{\alpha_0}{}^{m_0}{}_{\alpha_1}{}^{m_1}\cdots
_{\alpha_n}{}^{m_{n}}{}_{\alpha_{n+1}}(q_i)$ multiplied by
the one-loop factors from the vector and hypermultiplets.
The relation for the $n$-point function on a torus can be written down
in a similar manner.

The rewriting of the product of the DOZZ three-point functions
as the one-loop factor is analogous to what was presented above
for $\SU(2)$ $N_f=4$ theory and $\cN=2^*$ $SU(2)$ theory, so we
just mention two salient points.
Consider a three-point function $C(Q-\alpha_1,m,\alpha_2)$.
The denominator of the DOZZ formula then gives \begin{equation}
\prod \Gamma_2(\pm a_1\pm a_2 \pm \hat m +\frac Q2 )
\end{equation} where $\alpha_i=Q/2+a_i$, $m=Q/2+\hat m$.
This is the absolute value squared of the one-loop contribution
of a bifundamental hypermultiplet of mass $m$,
charged under two $\SU(2)$
gauge groups, with vevs $\pm a_1$  and $\pm a_2$ respectively.
Next consider the numerator of the DOZZ formula.
When we glue two three-point functions
along the channel where $V_\alpha$ is inserted,
we have the product of the form $C(\bullet ,\bullet ,\alpha)C(Q-\alpha , \bullet ,\bullet)$.
Then the numerator of the DOZZ formula gives \begin{align}
\Upsilon(2(Q-\alpha)) \Upsilon (2\alpha) & =
\left[\Gamma_2(2a)\Gamma_2(2a+Q)\Gamma_2(-2a)\Gamma_2(-2a+Q)\right]^{-1}\\
&=- 4a^2 \prod_{i=1,2}\left[ \Gamma_2(2a+\epsilon_i)\Gamma_2(-2a+\epsilon_i)\right]^{-1}
\end{align}
where we used the formula \eqref{gamma-shift}.
This gives the absolute value squared of the contribution
from the $\SU(2)$ vector multiplet with the vev $a$,
and also provides the crucial Vandermonde factor $a^2$.

\section{Seiberg-Witten differential and the insertion of $T(z)$}\label{sec:T}
Conformal blocks are only a fragment of a full CFT correlation function,
but, almost by construction, satisfy an important property: Ward identities for
the insertion of energy momentum tensor operators. This insertion  can be defined directly
by inserting the power expansion of the operator
\begin{equation}
T(z) = \sum L_{n} z^{-n-2}
\end{equation}
anywhere in the definition of the conformal block, or can be simply computed
through the Ward identity. On the sphere, for example,
\begin{equation}
\langle T(z) \prod_i {\cal O}_i(z_i) \rangle = \sum_j \left[ \frac{h_j}{(z-z_j)^2} + \frac{\partial_j}{z-z_j} \right]\langle \prod_i {\cal O}_i(z_i) \rangle
\end{equation}
where the insertion is made at the level of the conformal block. Do such energy momentum tensor insertions have any interesting meaning in the gauge theory side?
We can define a useful quadratic differential
\begin{equation}
\phi_2(z) dz^2 = -\frac{\langle T(z) \prod_i {\cal O}_i(z_i) \rangle }{\langle \prod_i {\cal O}_i(z_i) \rangle} \label{ward}
\end{equation}
$\phi_2(z)$ has double poles at $z_i$ with coefficient $-h_i$. The space of quadratic differentials
with double poles of fixed coefficients is an affine space of dimension $3g-3+n$. This is also the dimension of the Coulomb branch of ${\cal T}_{g,n}$. Indeed the Seiberg-Witten curve of the theory
can be also written as a double cover of $C_{g,n}$,
 in terms of a quadratic differential $\phi_2^{SW}(z)$, as
\begin{equation}
x^2 = \phi_2^{SW}(z)
\end{equation}

The coefficients of the double poles of $\phi_2^{SW}(z)$ are the squared mass parameters $-m_i^2$, i.e. \begin{equation}
m_a=\frac{1}{2\pi i}\oint_{\beta_a} xdz \label{mint}
\end{equation} where $\beta_a$ is a small circle around the $a$-th puncture.
The other $3g-3+n$ moduli can be fixed in terms of the special coordinates $a_i$ by computing the electric periods of the Seiberg-Witten differential
\begin{equation}
a_i = \frac{1}{2\pi i}\oint_{\gamma_i} x dz\label{aint}
\end{equation}
The cycles $\gamma_i$ are defined at weak coupling as wrapping around the long $i$-th tube.

The Seiberg-Witten curve is supposed to emerge from the Nekrasov partition function in the ``semiclassical limit'' $\epsilon_{1,2} \ll  a_i,m_i$.
We expect that
\begin{equation}
\phi_2(z) \to \phi_2^{SW}(z)
\end{equation}   in the same limit.
The property \eqref{mint}
can be easily checked:  we have
\begin{equation}
\frac{1}{2\pi i}\oint_{\beta_a} \sqrt{\phi_2(z)} = \sqrt{-h_a} \to m_a.
\end{equation}
Here we used \eqref{ward} in the first equality, and 
$h_a=m_a(Q-m_a)$ where $Q=\epsilon_1+\epsilon_2$ in the second limit.
We have also checked that \begin{equation}
\frac{1}{2\pi i}\oint_{\gamma_i} \sqrt{\phi_2(z)} \to a_i 
\end{equation} 
to high order in the expansion of the conformal blocks for ${\cal T}_{0,4}$, $\mathcal{T}_{0,5}$, ${\cal T}_{1,0}$  and ${\cal T}_{1,0}$. 
The agreement is quite remarkable, as the coefficients of $\phi_2$ and $\phi_2^{SW}$ are very intricate functions of $a_i, m_i, \tau_i$.
We expect this to be true for all $g,n$.

We are then led to speculate that at finite $\epsilon_{1,2}$, the notion of Seiberg-Witten curve should be
``quantized'' to the operator equation $x^2+T(z)=0$.

\section{Conclusions and Open Problems}\label{sec:speculation}
In this paper we considered Nekrasov's partition function
of four-dimensional $\cN=2$ theories which arise from
compactification of six-dimensional (2,0) theory of type $A_1$
on a sphere or a torus with punctures.
We showed that the instanton part of the partition function
gives the conformal blocks, and the one-loop part
gives the products of the DOZZ three-point functions of the Liouville theory.
Therefore, the integral of Nekrasov's full partition function
over the vevs of the adjoint scalars gives the Liouville correlation functions.

With these observations at hand, we propose the following general statement:
Given a genus-$g$ Riemann surface with $n$ punctures and
a particular sewing of the surface from three-punctured spheres,
consider the generalized quiver gauge theory naturally associated to it.
Then, {\em the conformal block for this sewing
is the instanton part of Nekrasov's partition function of this gauge theory}.
Furthermore,  {\em the $n$-point function of the Liouville theory
on this Riemann surface is equal to the integral 
of the absolute value squared of Nekrasov's full
 partition function of this gauge theory.}
The dictionary between the two sides of the correspondence is summarized in Table~\ref{table}.

\begin{table}[h]
\centering
\begin{tabular}{|@{$\Bigm|$}c|c@{$\Bigm|$}|}
\hline
\textbf{Gauge theory}  & \textbf{Liouville theory}\\
\hline
\hline
  & Liouville parameters \\
Deformation parameters $\epsilon_1$, $\epsilon_2$ & $\epsilon_1:\epsilon_2=b:1/b$ \\
& $c=1+6Q^2$, $Q=b+1/b$ \\
\hline
four free hypermultiplets & a three-punctured sphere \\
\hline
Mass parameter $m$   & Insertion of  \\
associated to an $\SU(2)$ flavor &  a Liouville exponential $ e^ {2m\phi}$ \\
\hline
one $\SU(2)$ gauge group & a thin neck (or channel)  \\
with UV coupling $\tau$ & with sewing parameter $q=\exp (2\pi i \tau)$ \\
\hline
Vacuum expectation value $a$  &  Primary $e^{2\alpha\phi} $ for the channel, \\
of an $\SU(2)$ gauge group & $\alpha=Q/2+a$  \\
\hline
Instanton part of $Z$ & Conformal blocks \\
\hline
One-loop part of $Z$ & Product of DOZZ factors \\
\hline
Integral of $|Z_\text{full}^2|$ & Liouville correlator \\
\hline
\hline
\end{tabular}
\caption{Dictionary between the Liouville correlation functions
and Nekrasov's partition function $Z$.\label{table} }
\end{table}


There are many open problems which beg to be answered.
We list them in a random order:
\begin{enumerate}
\item Prove mathematically the relation between
conformal blocks and
Nekrasov's partition function of the $U(2)$ quiver theory,
stripped of the $U(1)$ part.  Techniques developed in \cite{Eguchi:2003it,Iqbal:2004ne} might be useful.
\item Understand better the $U(1)$ part which we stripped manually.
\item Our identification of Nekrasov's instanton partition function
and the conformal block tells us how the former  transforms
under the S-duality,
in terms of the crossing symmetry of the conformal block.  
Find a physical explanation of this transformation law. cf.~\cite{Witten:1993ed}.
\item What is the relation of the Liouville theory and the theory
of chiral bosons ubiquitous in the topological vertex? cf.~\cite{Aganagic:2003db,Aganagic:2003qj}.
\item We found that Liouville correlators are the integral of the absolute value squared
of Nekrasov's full partition function. This should be  related to the OSV conjecture \cite{Ooguri:2004zv}. Make the relation precise.
\item Calculate directly Nekrasov's partition function for $SU(2)$ quivers,
rather than for $U(2)$ quivers. This should be possible by treating $SU(2)$
as either $SO(3)$ or as $Sp(1)$, cf.~\cite{Marino:2004cn,Nekrasov:2004vw}
\item Obtain Nekrasov's partition function when the gauge group
$\SU(2)_1\times\SU(2)_2\times \SU(2)_3$
couples to a half-hypermultiplet transforming in $\mathbf{2}_1\times \mathbf{2}_2
\times\mathbf{2}_3$.
This would be possible only in the $Sp(1)$ formulation.
Then compare the result with the appropriate conformal blocks.
\item Compactification on $S^4$  only gave us $b=1$.
Find manifolds which give $b\ne 1$.
\item We used Nekrasov's partition function to associate a number
to a gauge theory. We can also associate a number by taking the
partition function of the topologically-twisted gauge theory on K3 or
other four-manifolds, i.e.~by considering the Donaldoson invariants.
We would naively expect  to find a different  2d CFT
for each four-manifold. What are they?
\item Understand why we found Liouville theory, and why $\epsilon_1:\epsilon_2=b:1/b$. Our observation should have a place in the web of string dualities.
\item We interpreted the insertion of one Liouville energy-momentum tensor $T(z)$
as giving the Seiberg-Witten curve. What does multiple insertions of $T(z)$ correspond to?
We identified the correlators of Liouville primaries as the integral of  Nekrasov's partition function. What does the correlator of descendants correspond to?
\item Mathematically prove the equivalence of the square of the Seiberg-Witten
differential and the semi-classical limit of $\vev{T(z)}$. Combined with the proof
of the equivalence of the  instanton partition function and the conformal blocks,
this will provide a microscopic derivation of the Seiberg-Witten curve for
 quiver theories of $\SU(2)$ gauge groups. cf.~\cite{Shadchin:2005cc}.
\item It is natural in the  framework of \cite{Pestun:2007rz} to insert Wilson loops
on the gauge theory side.  What does it correspond to on the Liouville theory?
\item On the Liouville theory side, there are ZZ branes and FZZT branes.
What do they correspond to on the gauge theory side?
\item Both the Liouville theory  and the supersymmetric gauge theory
have been used to discuss geometric Langlands duality.
Does our observation have anything to say about it?
\item Extend the whole of our analysis to the $A_{N-1}$ theory,
i.e.~when $N$, not just two, M5-branes are used (and to type $D$ or $E$ as well).
The general structure of the instanton partition function appears to be compatible with the conformal blocks of W-algebras with a similar ADE classification.  The dimension $k$ currents should map to the degree $k$ differentials
in the canonical Seiberg-Witten  curve. The main problem is to relate the spectrum of possible punctures to the spectrum of highest weight representations of the W-algebras. The full partition function should be related to correlation functions of an ADE affine Toda theory,
a variant of Liouville theory which has  currents 
forming an ADE W-algebra.
\end{enumerate}
Solutions to any of the problems listed above would be welcomed.

\section*{Acknowledgments}
The authors have benefited from discussions with J. Maldacena, V. Pestun, N. Seiberg, C. Thorn and H. Verlinde. They also thank J. Teschner for informative correspondences.
L.F.A. and D.G. are supported in part by the DOE grant DE-FG02-
90ER40542. D.G. is supported in part by the Roger Dashen membership in the Institute for Advanced
Study. YT is supported in part by the NSF grant PHY-0503584, and by the Marvin L.
Goldberger membership at the Institute for Advanced Study.

\appendix

\section{Liouville theory}\label{app:liouville}
Here rudimentary facts of the Liouville theory are collected.
More details can be found in the reviews \cite{Teschner:2001rv,Nakayama:2004vk}.

\subsection{Conformal blocks}

Conformal blocks for the Liouville theory are just the ones of the Virasoro algebra.
These can be computed by the sewing procedure \cite{Belavin:1984vu,Sonoda:1988mf,Moore:1988qv,DiFrancesco:1997nk}. The basis for this procedure is the following schematic expression

\begin{equation}
\label{cb}
\langle {\cal O}_1 \cdots {\cal O}_k {\cal O}_{k+1}\cdots {\cal O}_n \rangle =\sum_{i,{\bf m},\bf{n}}  \langle {\cal O}_1 \cdots {\cal O}_k {\cal L}_{-\bf{m}} \phi_i  \rangle_M K^{-1}_{MN}  \langle {\cal L}_{-\bf{n}}  \phi_i  {\cal O}_{k+1}\cdots {\cal O}_n \rangle_N
\end{equation}
which allows to compute conformal blocks of a certain order in terms of lower order conformal blocks. The index $i$ runs over the primary fields and $\bf{m,n}$ run over the descendants of such primary fields, namely

\begin{equation}
{\cal L}_{-\bf{n}}  \phi_i =L_{-n_1}L_{-n_2}\cdots L_{-n_N}\phi_i
\end{equation}
where $L_n$ are the generators of the Virasoro algebra with central charge $c$, and $k=\sum_{i=1}^N n_i$ is the level of the descendant. The matrix $K$ is the so called Gram matrix, whose determinant is the Kac determinant. At level $k$, the indexes $M,N$ run over  the partitions of $k$, and  $K_{MN}$ are given by the inner product of the corresponding descendants of the primary field under consideration. For instance, at level two we get

\begin{equation}
K=\begin{pmatrix} \langle h|L_2 L_{-2}|h \rangle & \langle h|L_1^2 L_{-2}|h \rangle \cr \langle h|L_2 L_{-1}^2|h \rangle & \langle h|L_1^2 L_{-1}^2|h \rangle \end{pmatrix} =\begin{pmatrix} 4h+c/2 & 6h \cr 6h & 4h(1+2h) \end{pmatrix}
\end{equation}
where $|h \rangle$ is a primary of dimension $h$. It is then possible to express the conformal blocks we use in this paper
by sewing elementary building blocks, of the form
\begin{eqnarray}
R_M(h_1,h_2,h_3)= \frac{\langle {\cal O}_1{\cal O}_2 {\cal L}_{-\bf{m}} \phi_3  \rangle}{\langle {\cal O}_1{\cal O}_2 \phi_3  \rangle} \cr
S_{M,N}(h_1,h_2,h_3)= \frac{\langle {\cal L}_{-\bf{m}}  \phi_1 {\cal O}_2 {\cal L}_{-\bf{n}} \phi_3  \rangle}{\langle \phi_1 {\cal O}_2 \phi_3  \rangle}
\end{eqnarray}

\begin{figure}[ht]
\centering
\includegraphics[scale=0.4]{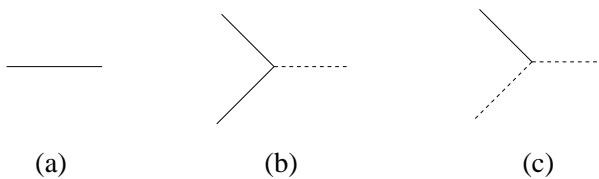}
\caption{Pictorial representation of the propagator $K^{-1}$, shown in (a), the vertex with two external legs $R$, shown in (b) and the vertex with one external leg $S$, shown in (c). External legs are represented with solid lines, while internal legs are represented with dashed lines.} \label{fig:1}
\end{figure}

We can represent pictorially $R$ and $S$ as stars with three legs. The difference between the two is the amount of external legs, two for the former and one for the later. $K^{-1}$ then represents a propagator, which joins internal legs, see Figure~\ref{fig:1}.

$R$ and $S$ depend on the dimensions of the primaries under consideration, while $K$ depends on the dimension of the primary interchanged and on the central charge.

\begin{figure}[h]
\centering
\includegraphics[scale=0.3]{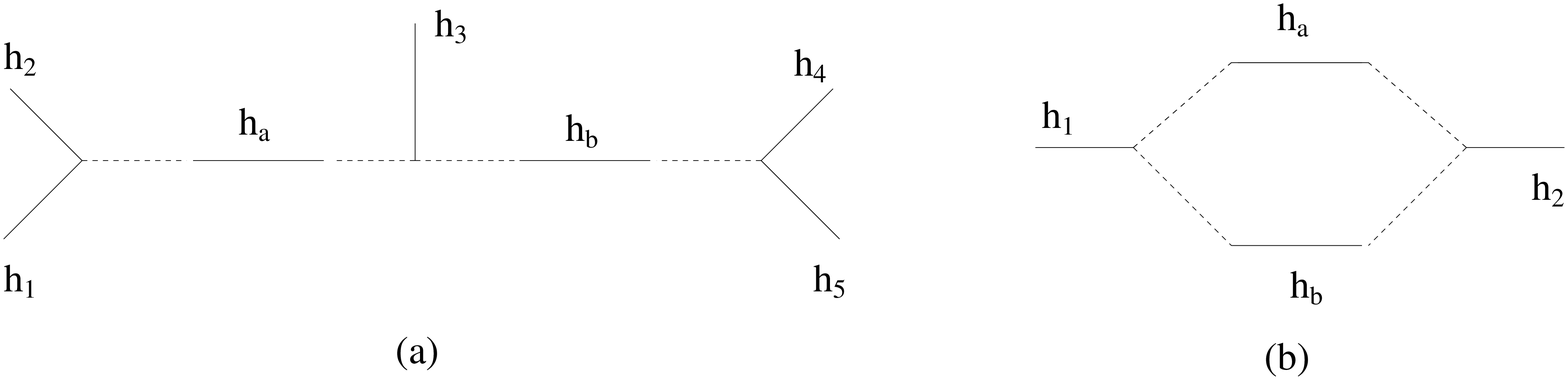}
\caption{Sewing of building blocks into sphere (a) and and torus (b) conformal blocks. In the figure we see a five-point conformal block in the sphere and two points on the torus.} \label{fig:2}
\end{figure}

Higher order conformal blocks can then be computed by sewing this building blocks, for instance, the four-point conformal block on the sphere is
\begin{equation}
R(h_4,h_3,h) K^{-1}(h) R(h,h_2,h_1),
\end{equation}
while that for five points is
\begin{equation}
R(h_4,h_5,h_b) K^{-1}(h_b) S(h_b,h_3,h_a)K^{-1}(h_a)R(h_a,h_2,h_1)
\end{equation}
and so on, while the one point on the torus is given by
\begin{equation}
\Tr \left(K^{-1}(h) S(h,h_1,h) \right),
\end{equation}
the two-point function is
\begin{equation}
\Tr \left(K^{-1}(h_a) S(h_a,h_2,h_b) K^{-1}(h_b)S(h_b,h_1,h_a)\right)
\end{equation}
 and so on, see Figure \ref{fig:2}.

The level of each contribution is then fixed by the level of its internal propagators, $k_1,k_2,\ldots,k_n$. The full conformal block is obtained by multiplying each contribution by $q_1^{k_1}\cdots q_n^{k_n}$ and adding up all contributions. For instance, for the five-point conformal block on the sphere and the two-point conformal block on the torus we obtain

\begin{eqnarray}
{\cal F}^{5pt}_{g=0}&=&1+\frac{(-h_1+h_2+h_a)(h_3+h_a-h_b)}{2h_a}q_1+\frac{(-h_4+h_5+h_b)(h_3-h_a+h_b)}{2h_b}q_2+\cdots \nonumber \cr
{\cal F}^{2pt}_{g=1}&=&1+\frac{(h_1+h_a-h_b)(h_2+h_a-h_b)}{2h_a}q_1+\frac{(h_1-h_a+h_b)(h_2-h_a+h_b)}{2h_b}q_2+\cdots  \nonumber
\end{eqnarray}

In section five we were interested in computing conformal blocks with an insertion of the energy momentum tensor $T(z)$. These can be easily computed by considering modified vertices, analogous to the ones defined above, which take into account the appropriate $T(z)$ insertion, for instance

\begin{eqnarray}
R \rightarrow RT= \frac{\langle {\cal O}_1{\cal O}_2 T(z){\cal L}_{-\bf{m}} \phi_3  \rangle}{\langle {\cal O}_1{\cal O}_2 \phi_3  \rangle} \cr
\end{eqnarray}
and so on, where $T(z)=\sum L_n z^{-n-2}$. Note that after inserting $T(z)$ even the conformal block at level zero is non trivial.

\subsection{The DOZZ formula}

Let us first define Barnes' double Gamma function \cite{Barnes}
which is ubiquitous in our discussion of the Liouville theory
and Nekrasov's partition function.
Barnes' double zeta function is \begin{equation}
\zeta_2(s;x|\epsilon_1,\epsilon_2)=\sum_{m,n} (m\epsilon_1+n\epsilon_2+x)^{-s}
=\frac{1}{\Gamma(s)} \int_0^\infty \frac{dt}{t} t^s
\frac{e^{-tx}}{(1-e^{-\epsilon_1 t})(1-e^{-\epsilon_2 t})}.
\end{equation}
This is the logarithm of Barnes' double-Gamma function,
\begin{equation}
\Gamma_2(x|\epsilon_1,\epsilon_2)=
\exp \frac{d}{ds}\Bigm|_0\zeta_2(s,x|\epsilon_1,\epsilon_2).
\end{equation} The arguments $\epsilon_{1,2}$ in $\Gamma_2$ will be
often omitted if there is no confusion.

Assume $\epsilon_{1,2}\in \mathbb{R}_{>0}$.
Then Barnes' double-Gamma function
is analytic in $x$ except at the poles at $x=-(m\epsilon_1+n\epsilon_2)$
where $(m,n)$ is a pair of non-negative integers.
Therefore one can think of Barnes' double-Gamma as the 
regularized infinite product
\begin{equation}
\Gamma_2(x|\epsilon_1,\epsilon_2) \propto \prod_{m,n \ge 0}
\left(x+m\epsilon_1+n \epsilon_2\right)^{-1}.\label{infiniteproduct}
\end{equation}
Furthermore it is real when $x$ is real.
As such, \begin{equation}
\Gamma_2(x^*)=
\Gamma_2(x)^*.
\end{equation}
Another relation we need is \begin{equation}
\Gamma_2(x+\epsilon_1)\Gamma_2(x+\epsilon_2)
=x \Gamma_2(x)\Gamma_2(x+\epsilon_1+\epsilon_2).\label{gamma-shift}
\end{equation}
This is a natural property the infinite product in the right hand side
of \eqref{infiniteproduct} would have.

We will also need the infinite product expansion when $\epsilon_1>0$, $\epsilon_2<0$,
which is given by \begin{equation}
\Gamma_2(x|\epsilon_1,\epsilon_2) \propto \prod_{m,n \ge 1}
\left(x+(m-1)\epsilon_1- n\epsilon_2\right)^{+1}.\label{infiniteproduct2}
\end{equation} Note that we have zeros instead of poles in this case.

The Liouville theory has the parameter $b$. The central chrage is then
\begin{equation}
c=1+6Q^2,\qquad Q=b+1/b.
\end{equation}
The three-point function 
is given by the DOZZ formula \cite{Dorn:1994xn,Zamolodchikov:1995aa,Teschner:1995yf,Teschner:2001rv,Teschner:2003en}
\begin{multline}
\vev{V_{\alpha_1}(z_1)V_{\alpha_2}(z_2)V_{\alpha_3}(z_3)}\\
=|z_{12}|^{2(\Delta_1+\Delta_2-\Delta_3)}|z_{23}|^{2(\Delta_2+\Delta_3-\Delta_1)}|z_{31}|^{2(\Delta_3+\Delta_1-\Delta_2)}
C(\alpha_1,\alpha_2,\alpha_3)
\end{multline}
where $\Delta_{i}$ is the dimension of the operators 
$V_{\alpha_i}=e^{2\alpha\phi}$ given by
\begin{equation}
\Delta_i=\alpha_i(Q-\alpha_i),
\end{equation}
and
\begin{multline}
C(\alpha_1,\alpha_2,\alpha_3)=\left[\pi \mu \gamma(b^2) b^{2-2b^2}\right]^{(Q-\alpha_1-\alpha_2-\alpha_3)/b} \\
\times \frac{\Upsilon'(0)\Upsilon(2\alpha_1)\Upsilon(2\alpha_2)\Upsilon(2\alpha_3)}
{\Upsilon(\alpha_1+\alpha_2+\alpha_3-Q)
\Upsilon(\alpha_1+\alpha_2-\alpha_3)
\Upsilon(\alpha_1-\alpha_2+\alpha_3)
\Upsilon(-\alpha_1+\alpha_2+\alpha_3)} \label{DOZZ}
\end{multline} where \begin{equation}
\Upsilon(x)=\frac{1}{\Gamma_2(x|b,b^{-1})\Gamma_2(Q-x|b,b^{-1})}.
\end{equation} and \begin{equation}
\gamma(x)= \Gamma(x)/\Gamma(1-x).
\end{equation}

\section{Nekrasov formulae}\label{app:nekrasov}

Here we provide the precise formulae of  Nekrasov's partition function\cite{Nekrasov:2002qd,Nekrasov:2003rj}
for the $U(2)$ quiver theories we discussed in the main part of the paper. 

\subsection{Instanton part}
The instanton partition function is
computed by performing the path integral by localizing with respect
to the $SO(4)$ rotation specified by $(\epsilon_1,\epsilon_2)$.
We set $\epsilon_+=\epsilon_1+\epsilon_2$; this was identified with $Q$
in the main part  of the text.
Localization fixes the gauge field configuration to be
instantons sitting at the origin; for each $U(2)$ gauge group
such fixed instantons are labeled by a pair of Young tableaux $\vec Y=(Y_1,Y_2)$,
and the instanton number is given by the total number  of boxes
$|\vec Y|=|Y_1|+|Y_2|$.  Then the contribution is weighted by the instanton factor
$q^{|\vec Y|}$ where \begin{equation}
q=\exp(2\pi i \tau_\text{UV}), \qquad
\tau_\text{UV}=\frac{4\pi i}{g^2_\text{UV}}+\frac{\theta_\text{UV}}{2\pi}.
\end{equation}
Note that the coupling constant receives finite renormalization,
even though the conformal quivers are finite theories; therefore it is important
to keep in mind that $\tau$ appearing in Nekrasov's partition function
is the UV coupling in a particular renormalization scheme.
We have seen that this scheme is a particularly natural one in the main part of the paper.

For a linear quiver with $N$ $U(2)$ gauge groups,
the partition function is then\begin{multline}
Z_\text{inst}= \sum_{\vec Y_1, \vec Y_2,\ldots, \vec Y_N}
\left(\prod_{i=1}^N q_i^{|\vec Y_i|} z_\text{vector}(\vec a_i,\vec Y_i) \right)
z_\text{antifund}(\vec a_1,\vec Y_1,\mu_1)
z_\text{antifund}(\vec a_1,\vec Y_1,\mu_2)  \\
\times \left(\prod_{i=1}^{N-1} z_\text{bifund}(\vec a_i,\vec Y_i;\vec a_{i+1},\vec Y_{i+1};m_i) \right)
z_\text{fund}(\vec a_N,\vec Y_N,\mu_3)
z_\text{fund}(\vec a_N,\vec Y_N,\mu_4).
\end{multline}
Here $\vec a_i=(a_{i,1},a_{i,2})$
is the diagonal of the adjoint scalar,
$\vec Y_i$  the pair of the Young tableaux specifying the fixed instanton,
$q_i$ is the exponentiated
UV gauge coupling of the $i$-th $SU(2)$ gauge group.
$m_i$ is the mass of the bifundamental hypermultiplet charged under $SU(2)_i$
and $SU(2)_{i+1}$. $\mu_{1,2,3,4}$ are the masses of the fundamentals.
$z_\text{vector}$, $z_\text{bifund}$ etc. are the contribution
of the vector multiplet,
the bifundamental hypermultiplet, etc. defined below.
For a necklace quiver one needs to replace four fundamentals
by one bifundamental charged under $SU(2)_n$ and $SU(2)_1$.
For the $\mathcal{N}=2^*$ $SU(2)$  theory one needs to put the contribution
from the adjoint hypermultiplet $z_\text{adj}$.

\begin{figure}
\centerline{\includegraphics[scale=.3]{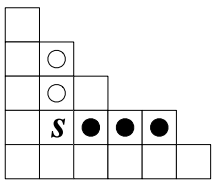}}
\caption{Definition of the arm-length and the leg-length. For a box $s$ in a Young tableau displayed above, the leg-length is the number of boxes to the right of $s$,
marked by black disks, and the arm-length is the number of boxes on top of $s$.\label{fig:hook}}
\end{figure}

Let $Y=(\lambda_1\ge \lambda_2\ge \cdots)$ be a Young tableau
where $\lambda_i$ is the height of the $i$-th column.
We set $\lambda_i=0$ when $i$ is larger than the width of the tableau.
Let $Y^T=(\lambda_1'\ge \lambda_2\ge \cdots)$ be its transpose.
For a box $s$ at the coordinate $(i,j)$, we let its arm-length $A_Y(s)$
and leg-length $L_Y(s)$ with respect to the tableau $Y$ to be \begin{equation}
A_Y(s)=\lambda_i-j,\qquad
L_Y(s)=\lambda_j'-i,
\end{equation}  see Fig.~\ref{fig:hook}. Note that they can be negative when
the box $s$ is outside the tableau.
We then define a function $E$ by \begin{equation}
E(a,Y_1,Y_2,s)=a-\epsilon_1 L_{Y_2}(s) + \epsilon_2 (A_{Y_1}(s)+1).
\end{equation}

We use the vector symbol $\vec a$  to stand for pairs $\vec a =(a_1,a_2)$,
e.g.~$\vec a_1=(a_{1,1},a_{2,1})$, $\vec Y=(Y_1,Y_2)$, etc.
We can now define the contribution of a bifundamental \cite{Fucito:2004gi,Shadchin:2005cc}: \begin{multline}
z_\text{bifund}(\vec a,\vec Y;\vec b,\vec W;m)= \\
\prod_{i,j=1}^2
\prod_{s\in Y_i}( E(a_i-b_j,Y_i,W_j,s)- m)
\prod_{t\in W_j}(\epsilon_+  -  E(b_j-a_i,W_j,Y_i,t)- m)
\end{multline}
Note that this is {\em not} symmetric under the exchange
between $(\vec a,\vec Y)$ and $(\vec b,\vec W)$; instead it satisfies \begin{equation}
z_\text{bifund}(\vec a,\vec Y;\vec b,\vec W;m)=
z_\text{bifund}(\vec b,\vec W;\vec a,\vec Y;\epsilon_+-m).
\end{equation}
This behavior comes from a subtlety in Nekrasov's setup:
 the gauge group is strictly speaking $U(2)\times U(2)$ instead of
$SU(2)\times SU(2)$, so the representations $\mathbf{2}\otimes \bar{\mathbf{2}}$
and $\bar{\mathbf{2}}\otimes \mathbf{2}$ are different.
The formula above means that by exchanging a bifundamental with
a anti-bifundamental, we need to flip the mass as $m\to \epsilon_+-m$.

The contribution of an adjoint hypermultiplet is now easy to define.
It is \begin{equation}
z_\text{adj}(\vec a,\vec Y,m)=z_\text{bifund}(\vec a,\vec Y,\vec a,\vec Y,m).
\end{equation}
Then the contribution of a vector multiplet is
\begin{equation}
z_\text{vector}(\vec a,\vec Y)=1/z_\text{adj}(\vec a,\vec Y,0).
\end{equation}

The contribution from (anti)fundamental hypermultiplets is defined as follows:
\begin{align}
z_\text{fund}(\vec a,\vec Y,m)&=\prod_{i=1}^2 \prod_{s\in Y_i} ( \phi(a_i,s) -m+\epsilon_+), \\
z_\text{antifund}(\vec a,\vec Y,m)&=
z_\text{fund}(\vec a,\vec Y,\epsilon_+-m)
\end{align} where $\phi(a,s)$ for the box $s=(i,j)$ is defined as \begin{equation}
\phi(a,s)=a+\epsilon_1 (i-1)+\epsilon_2(j-1).
\end{equation}
$z_\text{(anti)fund}$ and $z_\text{bifund}$ satisfies two important relations:
\begin{align}
z_\text{bifund}(\vec a,\vec Y,\vec \mu, \emptyset,m)
&= z_\text{fund}(\vec a,\vec Y,m+\mu)
z_\text{fund}(\vec a,\vec Y, m-\mu),\label{bifund-vs-fund1}\\
z_\text{bifund}(\vec \mu, \emptyset,\vec a,\vec Y,m)
&= z_\text{antifund}(\vec a,\vec Y,m+\mu)
z_\text{antifund}(\vec a,\vec Y, m-\mu)\label{bifund-vs-fund2}
\end{align}
where  $\vec \mu=(\mu_1,\mu_2)=(\mu,-\mu)$, and $\emptyset$ stands
for a pair of empty Young tableaux.
The relation \eqref{bifund-vs-fund1}
means that a bifundamental, with the second $SU(2)$ at zero coupling,
behaves as two fundamental hypermultiplets of the first $SU(2)$ with mass $m\pm \mu$
where $(\mu,-\mu)$ are  the  diagonal entries of the adjoint scalar of the second $SU(2)$;
this is as expected. The relation \eqref{bifund-vs-fund2} can be understood similarly.

\subsection{One-loop part}\label{sec:one-loop}
The partition function discussed above contains only the contribution from
the instantons. The full partition function also contains the classical and the one-loop part,
i.e. \begin{equation}
Z_\text{Nekrasov}= Z_\text{classical} Z_\text{1-loop} Z_\text{inst}.
\end{equation} The prepotential $F$ can then be recovered from its logarithm,
\begin{equation}
F(\tau_i;m_a;a_i)= \lim_{\epsilon_1,\epsilon_2\to 0} \epsilon_1\epsilon_2
 \log Z_\text{Nekrasov} (\tau_i;m_i;a_i).
\end{equation} 

The classical part is simply \begin{equation}
Z_\text{classical}= \exp\left[-\frac{1}{\epsilon_1\epsilon_2}\sum_i (2\pi i)\tau_i a_i^2\right].
\end{equation}

The basic ingredient of the one-loop part is
the logarithm of Barnes' double gamma functions:
\begin{equation}
\gamma_{\epsilon_1,\epsilon_2}(x)=\log \Gamma_2(x+\epsilon_+|\epsilon_1,\epsilon_2).
\end{equation}

When $\hbar=\epsilon_1=-\epsilon_2$,
$\gamma_{\hbar,-\hbar}(x)$
has the expansion \begin{equation}
\gamma_{\hbar,-\hbar}(x) = \hbar^{-2}\left(
\frac12 x^2\log x - \frac 34 x^2
\right) - \frac 1{12}\log x +
\sum_{g=2}^\infty \frac{B_{2g}}{2g(2g-2)} \left(\frac{\hbar}{x}\right)^{2g-2}
\end{equation} where $B_{2g}$ is the Bernoulli number;
one recognizes the limit
\begin{equation}
\lim_{\hbar\to 0} \hbar^2 \gamma_{\hbar,-\hbar}(x) =
\frac12 x^2\log x - \frac 34 x^2
\end{equation} to be the standard one-loop contribution to the prepotential
of a hypermultiplet of mass $x$, i.e.~$[(x^2/2)\log x-(3/4)x^2]''=\log x$.

Then, for a linear quiver with $N$ gauge groups, the one-loop part is
given by combining this factor for all elementary particles:
\begin{multline}
Z_\text{1-loop}=
\left(\prod_{i=1}^N  z_\text{vector}^\text{1-loop}(\vec a_i) \right)
z_\text{antifund}^\text{1-loop}(\vec a_1,\mu_1)
z_\text{antifund}^\text{1-loop}(\vec a_1,\mu_2)  \\
\times \left(\prod_{i=1}^{N-1} z_\text{bifund}^\text{1-loop}(\vec a_i,\vec a_{i+1},m_i) \right)
z_\text{fund}^\text{1-loop}(\vec a_N,\mu_3)
z_\text{fund}^\text{1-loop}(\vec a_N,\mu_4)
\end{multline}
where \begin{align}
z_\text{vector}^\text{1-loop}(\vec a) &=
\prod_{i<j} \exp \left[-\gamma_{\epsilon_1,\epsilon_2} (a_i-a_j-\epsilon_1) -
\gamma_{\epsilon_1,\epsilon_2} (a_i-a_j-\epsilon_2)\right],\label{oneloop-vector}\\
z_\text{fund}^\text{1-loop}(\vec a,\mu) &=
\prod_{i} \exp \left[\gamma_{\epsilon_1,\epsilon_2} (a_i-\mu) \right],
\label{oneloop-hyper}\\
z_\text{antifund}^\text{1-loop}(\vec a,\mu) &=
\prod_{i} \exp \left[\gamma_{\epsilon_1,\epsilon_2} (-a_i+\mu-\epsilon_+) \right],\\
z_\text{bifund}^\text{1-loop}(\vec a,\vec b,m) &=
\prod_{i,j} \exp \left[\gamma_{\epsilon_1,\epsilon_2} (a_i-b_j-m) \right].
\end{align}
The contribution of a $U(N)$  adjoint hypermultiplet 
is given by $z_\text{bifund}^\text{1-loop}(\vec a,\vec a, m)$.

The one-loop factor for the vector multiplet \eqref{oneloop-vector}
is different from the one in \cite{Nekrasov:2003rj}, in that 
theirs does not have the shift by $-\epsilon_{1,2}$ in it.
A  more detailed analysis \cite{Hollowood:2003cv,Iqbal:2007ii} showed that \eqref{oneloop-vector}
is more appropriate. Using Eq.(128) of \cite{Hollowood:2003cv}
or using Eq.(3) of \cite{Iqbal:2007ii}, one finds that the contribution
of a five-dimensional vector multiplet 
with mass $\mu$, which has $j_R=1/2$ in their notation, 
is given by \begin{equation}
\prod_{m,n=1}^\infty
 (1-e^{-(m-1)\epsilon_1+n\epsilon_2-\mu})^{-1}
  (1-e^{-m\epsilon_1+(n-1)\epsilon_2-\mu})^{-1}.
\end{equation} 
Taking the four-dimensional limit  involves replacing $1-e^{-x}$ by $x$
in the infinite product. 
Here we need to remind ourselves $\epsilon_1>0$ and $\epsilon_2<0$.\footnote{The authors thank T. Okuda for pointing the error they made in v1.}
Using \eqref{infiniteproduct2}, we find that it becomes \begin{equation}
\Gamma_2(\mu|\epsilon_1,\epsilon_2)^{-1}
\Gamma_2(\mu+\epsilon_1+\epsilon_2|\epsilon_1,\epsilon_2)^{-1}
\end{equation}
which reproduces  \eqref{oneloop-vector}, up to one factor of $\mu$.

Similarly, using the same starting point but with $j_R=0$ which corresponds
to a hypermultiplet, one finds \begin{equation}
\prod_{m,n=1}^\infty
 (1-e^{-(m-1/2)\epsilon_1+(n-1/2)\epsilon_2-\mu})
\end{equation} which in the four-dimensional limit becomes \begin{equation}
\Gamma_2(\mu+\frac{\epsilon_+}2| \epsilon_1,\epsilon_2).
\end{equation} Again, we used \eqref{infiniteproduct2}.
This reproduces \eqref{oneloop-hyper} with the shift of $\mu$ by $\epsilon_+/2$.
It just reflects the difference in the conventions 
in \cite{Nekrasov:2003rj} and in \cite{Hollowood:2003cv,Iqbal:2007ii}
of the zero of the hypermultiplet mass.

\subsection{Example}\label{sec:renormalization}
As an example we present here a result
for  $SU(2)$ theory with $N_f=4$ massless flavors.
Performing the instanton sum
and combining with the one-loop and the classical parts,
we obtain the low-energy prepotential:\begin{multline}
(2\pi i)\tau_\text{IR} a^2=\\
(2\pi i)\tau_\text{UV} a^2
-(\log 16) a^2
+ a^2\left(
\frac12q+\frac{13}{64}q^2+\frac{23}{192}q^3+\frac{2701}{32768}q^4+\cdots
\right) \label{IR-UV-masslessNf4}
\end{multline} where $q=\exp(2\pi i \tau_\text{UV})$.
The first, the second, and the third terms come from
the classical, the one-loop and the instanton contributions, respectively.
Let us define \begin{equation}
q_\text{IR}=\exp(4\pi i \tau_\text{IR}), \quad
\tau_\text{IR}=\frac{4\pi i}{g^2_\text{IR}}+\frac{\theta_\text{IR}}{2\pi}.
\end{equation}
Note that in the presence of the massless hypermultiplets 
in the doublet representation
there is a shift symmetry of the theta angle, $\theta\to \theta+\pi$.
$q_\text{IR}$ is designed to be invariant under this shift.

Inverting \eqref{IR-UV-masslessNf4}, we find \begin{equation}
q_\text{UV}=16 q_\text{IR}^{1/2}-128 q_\text{IR}
+704q_\text{IR}^{3/2}-3072 q_\text{IR}^2+\cdots
=\frac{\theta_2(q_\text{IR})^4}{\theta_3(q_\text{IR})^4},
\end{equation} or equivalently $q_\text{UV}=\lambda(2\tau_\text{IR})$
where $\lambda$ is the modular lambda function.
This means that the double cover of a sphere with
four branch points with cross ratio $q_\text{UV}$
is an elliptic curve with modulus $2\tau_\text{IR}$.
This relation was first noticed in \cite{Grimm:2007tm}.

\section{$U(1)$ factors}\label{app:u1}
According to our conjecture, Nekrasov's instanton partition function on certain generalized quiver theories coincides with the conformal blocks on the corresponding two dimensional Riemann surface, upon stripping off a ``$U(1)$ factor'' which presumably arises from the fact that the Nekrasov partition function is computed for $U(2)$ groups, rather than $SU(2)$. In this appendix we write down the explicit form of such factors for the case of the sphere and the torus.

\subsection{Sphere}

\begin{figure}
\centering
\includegraphics[scale=0.3]{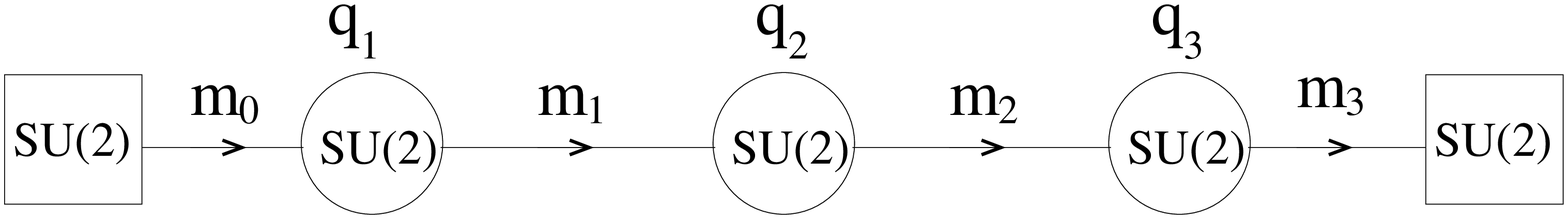}
\caption{Quiver theory corresponding to the sphere with six punctures.} \label{fig:quiver1}
\end{figure}

The quiver gauge theory corresponding to the sphere with $n+3$ punctures has $\prod_i^n SU(2)_i$ gauge group, as shown in Figure~\ref{fig:quiver1}, whose instanton numbers are counted by powers of $q_i$. In addition we have bi-fundamental matter with masses $m_i$, $i=0,\ldots,n$, and $m_0 \equiv m_{n+1}$. The explicit result for the first few cases is

\begin{eqnarray}
Z_{U(1)}^{g=0,n=4}&=&(1-q)^{2 m_0 (Q-m_1)}\\
Z_{U(1)}^{g=0,n=5}&=&(1-q_1)^{2 m_0 (Q-m_1)}(1-q_2)^{2 m_1 (Q-m_2)}
(1-q_1 q_2)^{2 m_0 (Q-m_2)}\\
Z_{U(1)}^{g=0,n=6}&=&(1-q_1)^{2 m_0 (Q-m_1)}(1-q_2)^{2 m_1 (Q-m_2)}
(1-q_3)^{2 m_2 (Q-m_3)} \times \\
\times & &(1-q_1 q_2)^{2 m_0 (Q-m_2)}(1-q_2 q_3)^{2 m_1 (Q-m_3)}(1-q_1 q_2 q_3)^{2 m_0 (Q-m_3)} \nonumber
\end{eqnarray}

Note that the general pattern is easily recognizable. Given a set of consecutive nodes, let's say $q_1,\ldots,q_m$, there is a factor of the form
\begin{equation}
\prod_{\ell=0}^\infty(1-q_1\cdots q_m)^{2m_\text{in}(Q-m_\text{out})},
\end{equation}
where $m_\text{in}$ is the mass of the adjoint bifundamental that enters into the set of nodes and $m_\text{out}$ is the mass of the adjoint bifundamental that exits the set of nodes. One then has to multiply the contributions from all such sets. 
We checked this general pattern up to $n=7$.

Note that when defining $m_{in}$ and $m_{out}$, we have assigned an orientation for the bifundamental matter, shown by the direction of the arrows in figure~\ref{fig:quiver1}. This of course wouldn't make sense when considering $SU(2)$ gauge groups, but remember that Nekrasov's partition function was computed for $U(2)$ quiver gauge theories.

\subsection{Torus}

\begin{figure}
\centering
\includegraphics[scale=0.3]{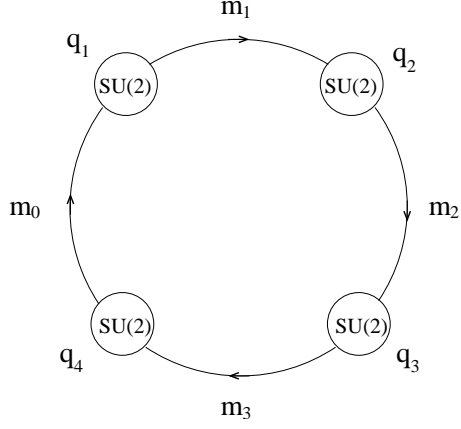}
\caption{Necklace quiver theory corresponding to the torus with four punctures.} \label{fig:quiver2}
\end{figure}

In the case of the torus, we are led to consider the instanton partition function of necklace quiver theories as the one shown in Figure~\ref{fig:quiver2}.
These necklace theories have $\prod_i^n SU(2)_i$ gauge group, as shown in the figure, whose instanton numbers are counted by powers of $q_i$. In addition, we have bi-fundamental matter with mass $m_0 \equiv m_{n}$, $m_1$, etc. Furthermore, when computing the instanton partition function we need to choose an orientation for the bi-fundamental matter, as explained above. 

The explicit result for the first few cases is
\begin{eqnarray}
Z_{U(1)}^{g=1,n=1}&=&\prod_{\ell=0}^\infty(1-q_1^{\ell+1})^{2m_1(Q-m_1)-1}  \\
Z_{U(1)}^{g=1,n=2}&=&\prod_{\ell=0}^\infty(1-q_1^{\ell+1}q_2^{\ell+1})^{2m_1(Q-m_1)+2m_2(Q-m_2)-1}\times \nonumber \\
\times& & (1-q_1^{\ell+1}q_2^{\ell})^{2 m_2(Q-m_1)}(1-q_1^{\ell}q_2^{\ell+1})^{2 m_1(Q-m_2)} \\
Z_{U(1)}^{g=1,n=3}&=&\prod_{\ell=0}^\infty(1-q_1^{\ell+1}q_2^{\ell+1}q_3^{\ell+1})^{2m_1(Q-m_1)+2m_2(Q-m_2)+2m_3(Q-m_3)-1} \times \nonumber \\
\times & &\text{\small $(1-q_1^{\ell+1}q_2^{\ell+1}q_3^{\ell})^{2m_3(Q-m_2)} (1-q_1^{\ell+1}q_2^{\ell}q_3^{\ell+1})^{2m_2(Q-m_1)} (1-q_1^{\ell}q_2^{\ell+1}q_3^{\ell+1})^{2m_1(Q-m_3)} \times$} \nonumber \\
\times & & \text{\small$ (1-q_1^{\ell+1}q_2^{\ell}q_3^{\ell})^{2m_3(Q-m_1)} (1-q_1^{\ell}q_2^{\ell+1}q_3^{\ell})^{2m_1(Q-m_2)} (1-q_1^{\ell}q_2^{\ell}q_3^{\ell+1})^{2m_2(Q-m_3)} $}
\end{eqnarray}

The general pattern is easy to recognize. First of all, given the quiver diagram with $n-$nodes, there is a factor
\begin{equation}
\prod_{\ell=0}^\infty(1-q_1^{\ell+1}\cdots q_n^{\ell+1})^{\sum_{i=1}^n 2m_i(Q-m_i)-1}.
\label{foo}
\end{equation}
Second, given a set of consecutive nodes, let's say $q_1,\ldots,q_m$, there is a factor of the form
\begin{equation}
\prod_{\ell=0}^\infty(1-q_1^{\ell+1}\cdots q_m^{\ell+1} q_{m+1}^\ell \cdots q_n^\ell)^{2m_\text{in}(Q-m_\text{out})},\label{bar}
\end{equation}
where $m_\text{in}$ is the mass of the adjoint bifundamental that enters into the set of nodes and $m_\text{out}$ is the mass of the adjoint bifundamental that exits the set of nodes.
Another interpretation is to consider all possible paths on the quiver
starting on a node, say $q_i$, and go $k$ steps along the arrow,
winding many times on the circle. Then we can easily see that
the expressions written above equals \begin{equation}
Z^{g=1}_{U(1)}=
\frac{\prod_{i=1}^n  \prod_{k=0}^\infty 
(1-q_i q_{i+1} \cdots q_{i+k})^{2m_{i-1} (Q-m_{i+k})}}
{
\prod_{\ell=1}^\infty(1-q_1^{\ell}q_2^{\ell}\cdots q_n^{\ell})
 }
\end{equation} where the subscripts of $q_i$ and $m_i$ are considered modulo $n$.
We checked this general pattern up to $n=4$.

As a further check, note that if we disconnect one of the nodes, sending $q_n \rightarrow 0$, we reobtain the $U(1)$ factors of the linear quiver (remembering that $m_n$ and $m_0$ are identified.)

\bibliography{All}{}
\end{document}